%%%%%%%%%%%%%%%%%%     PLAIN TEX FILE       %%%%%%%%%%%%%%%%%%
 %%%%%%%%%%%%%%%%%%  %%%%%%%%%%%%%%%%%%  %%%%%%%%%%%%%%%%%%  %%%%%%%%%%%%%%%%%%
 %%%%%%%%%%%%%%%%%%  %%%%%%%%%%%%%%%%%%  %%%%%%%%%%%%%%%%%%  %%%%%%%%%%%%%%%%%%
 %%%%%%%%%%%%%%%%%%  %%%%%%%%%%%%%%%%%%  %%%%%%%%%%%%%%%%%%  %%%%%%%%%%%%%%%%%%

 %%%%%%%%%%%%%%%%%%  tex macros for preprints, cm version %%%%%%%%%%%%%%
%                     (P. Ginsparg, last updated 9/91)
%                if confused, type `b' in response to query
%
%---------------------------------------------------------------------%
%% site dependent options:
%% \unredoffs and \redoffs define horizontal and vertical offsets
%% respectively for unreduced and reduced modes. \speclscape defines
%% the \special{} call that sets printer to landscape (sideways) mode.
%% from standard set below, leave uncommented as appropriate or redefine
%
%%% next 400dpi
%\def\unredoffs{} \def\redoffs{\voffset=-.31truein\hoffset=-.48truein}
%\def\speclscape{\special{landscape}}
%
%%% apple lw
\def\unredoffs{} \def\redoffs{\voffset=-.31truein\hoffset=-.59truein}
\def\speclscape{\special{ps: landscape}}
%
%%% qms lasergrafix:
%\def\unredoffs{} \def\redoffs{\voffset=-.4truein\hoffset=.125truein}
%\def\speclscape{\special{qms: landscape}}
%
%%% saclay A4 paper:
%\def\unredoffs{\hoffset-.14truein\voffset-.2truein}
%\def\redoffs{\voffset=-.55truein\hoffset=-.1truein} \def\speclscape{}
%
%---------------------------------------------------------------------%
%
\newbox\leftpage \newdimen\fullhsize \newdimen\hstitle \newdimen\hsbody
\tolerance=1000\hfuzz=2pt
\catcode`\@=11 % This allows us to modify PLAIN macros.
\def\bigans{b }
%\message{ big or little (b/l)? }\read-1 to\answ
\def\answ{b }
\ifx\answ\bigans\message{(This will come out unreduced.}
\magnification=1200\unredoffs\baselineskip=16pt plus 2pt minus 1pt
\hsbody=\hsize \hstitle=\hsize %take default values for unreduced format
\else\message{(This will be reduced.} \let\l@r=L
\magnification=1000\baselineskip=16pt plus 2pt minus 1pt \vsize=7truein
\redoffs \hstitle=8truein\hsbody=4.75truein\fullhsize=10truein\hsize=\hsbody
\output={\ifnum\pageno=0 %%% This is the HUTP version
  \shipout\vbox{\speclscape{\hsize\fullhsize\makeheadline}
    \hbox to \fullhsize{\hfill\pagebody\hfill}}\advancepageno
  \else
  \almostshipout{\leftline{\vbox{\pagebody\makefootline}}}\advancepageno
  \fi}
\def\almostshipout#1{\if L\l@r \count1=1 \message{[\the\count0.\the\count1]}
      \global\setbox\leftpage=#1 \global\let\l@r=R
 \else \count1=2
  \shipout\vbox{\speclscape{\hsize\fullhsize\makeheadline}
      \hbox to\fullhsize{\box\leftpage\hfil#1}}  \global\let\l@r=L\fi}
\fi
%---------------------------------------------------------------------
%
\newcount\yearltd\yearltd=\year\advance\yearltd by -1900

\def\Title#1#2{\nopagenumbers\abstractfont\hsize=\hstitle\rightline{#1}%
\vskip 1in\centerline{\titlefont #2}\abstractfont\vskip .5in\pageno=0}
\def\Date#1{\vfill\leftline{#1}\tenpoint\supereject\global\hsize=\hsbody%
\footline={\hss\tenrm\folio\hss}}%      restores pagenumbers
%
%       use following instead of \Date on the preliminary draft,
%       puts date/time on each page in big mode, writes labels in margins

\def\draftmode{\message{ DRAFTMODE }\def\draftdate{{\rm preliminary draft:
\number\month/\number\day/\number\yearltd\ \ \hourmin}}%
\headline={\hfil\draftdate}\writelabels\baselineskip=20pt plus 2pt minus 2pt
 {\count255=\time\divide\count255 by 60 \xdef\hourmin{\number\count255}
  \multiply\count255 by-60\advance\count255 by\time
  \xdef\hourmin{\hourmin:\ifnum\count255<10 0\fi\the\count255}}}
%       use \nolabels to get rid of eqn, ref, and fig labels in draft mode
\def\nolabels{\def\wrlabeL##1{}\def\eqlabeL##1{}\def\reflabeL##1{}}
\def\writelabels{\def\wrlabeL##1{\leavevmode\vadjust{\rlap{\smash%
{\line{{\escapechar=` \hfill\rlap{\sevenrm\hskip.03in\string##1}}}}}}}%
\def\eqlabeL##1{{\escapechar-1\rlap{\sevenrm\hskip.05in\string##1}}}%
\def\reflabeL##1{\noexpand\llap{\noexpand\sevenrm\string\string\string##1}}}
\nolabels
%
% tagged sec numbers
\global\newcount\secno \global\secno=0
\global\newcount\meqno \global\meqno=1
\def\newsec#1{\global\advance\secno by1\message{(\the\secno. #1)}
%\ifx\answ\bigans \vfill\eject \else \bigbreak\bigskip \fi  %if desired
\global\subsecno=0\eqnres@t\noindent{\bf\the\secno. #1}
\writetoca{{\secsym} {#1}}\par\nobreak\medskip\nobreak}
\def\eqnres@t{\xdef\secsym{\the\secno.}\global\meqno=1\bigbreak\bigskip}
\def\sequentialequations{\def\eqnres@t{\bigbreak}}\xdef\secsym{}
\global\newcount\subsecno \global\subsecno=0
\def\subsec#1{\global\advance\subsecno by1\message{(\secsym\the\subsecno.
#1)}
\ifnum\lastpenalty>9000\else\bigbreak\fi
\noindent{\it\secsym\the\subsecno. #1}\writetoca{\string\quad
{\secsym\the\subsecno.} {#1}}\par\nobreak\medskip\nobreak}
\def\appendix#1#2{\global\meqno=1\global\subsecno=0\xdef\secsym{\hbox{#1.}}
\bigbreak\bigskip\noindent{\bf Appendix #1. #2}\message{(#1. #2)}
\writetoca{Appendix {#1.} {#2}}\par\nobreak\medskip\nobreak}
%
%       \eqn\label{a+b=c}       gives displayed equation, numbered
%                               consecutively within sections.
%     \eqnn and \eqna define labels in advance (of eqalign?)
%
\def\eqnn#1{\xdef #1{(\secsym\the\meqno)}\writedef{#1\leftbracket#1}%
\global\advance\meqno by1\wrlabeL#1}
\def\eqna#1{\xdef #1##1{\hbox{$(\secsym\the\meqno##1)$}}
\writedef{#1\numbersign1\leftbracket#1{\numbersign1}}%
\global\advance\meqno by1\wrlabeL{#1$\{\}$}}
\def\eqn#1#2{\xdef #1{(\secsym\the\meqno)}\writedef{#1\leftbracket#1}%
\global\advance\meqno by1$$#2\eqno#1\eqlabeL#1$$}
%
%                            footnotes
\newskip\footskip\footskip14pt plus 1pt minus 1pt %sets footnote baselineskip
\def\footnotefont{\ninepoint}\def\f@t#1{\footnotefont #1\@foot}
\def\f@@t{\baselineskip\footskip\bgroup\footnotefont\aftergroup\@foot\let\next}
\setbox\strutbox=\hbox{\vrule height9.5pt depth4.5pt width0pt}
\global\newcount\ftno \global\ftno=0
\def\foot{\global\advance\ftno by1\footnote{$^{\the\ftno}$}}
%
%say \footend to put footnotes at end
%will cause problems if \ref used inside \foot, instead use \nref before
\newwrite\ftfile
\def\footend{\def\foot{\global\advance\ftno by1\chardef\wfile=\ftfile
$^{\the\ftno}$\ifnum\ftno=1\immediate\openout\ftfile=foots.tmp\fi%
\immediate\write\ftfile{\noexpand\smallskip%
\noexpand\item{f\the\ftno:\ }\pctsign}\findarg}%
\def\footatend{\vfill\eject\immediate\closeout\ftfile{\parindent=20pt
\centerline{\bf Footnotes}\nobreak\bigskip\input foots.tmp }}}
\def\footatend{}
%
%     \ref\label{text}
% generates a number, assigns it to \label, generates an entry.
% To list the refs on a separate page,  \listrefs
%
\global\newcount\refno \global\refno=1
\newwrite\rfile
\def\ref{[\the\refno]\nref}
\def\nref#1{\xdef#1{[\the\refno]}\writedef{#1\leftbracket#1}%
\ifnum\refno=1\immediate\openout\rfile=refs.tmp\fi
\global\advance\refno by1\chardef\wfile=\rfile\immediate
\write\rfile{\noexpand\item{#1\ }\reflabeL{#1\hskip.31in}\pctsign}\findarg}
%        horrible hack to sidestep tex \write limitation
\def\findarg#1#{\begingroup\obeylines\newlinechar=`\^^M\pass@rg}
{\obeylines\gdef\pass@rg#1{\writ@line\relax #1^^M\hbox{}^^M}%
\gdef\writ@line#1^^M{\expandafter\toks0\expandafter{\striprel@x #1}%
\edef\next{\the\toks0}\ifx\next\em@rk\let\next=\endgroup\else\ifx\next\empty%
\else\immediate\write\wfile{\the\toks0}\fi\let\next=\writ@line\fi\next\relax}}
\def\striprel@x#1{} \def\em@rk{\hbox{}}
\def\lref{\begingroup\obeylines\lr@f}
\def\lr@f#1#2{\gdef#1{\ref#1{#2}}\endgroup\unskip}
\def\semi{;\hfil\break}
\def\addref#1{\immediate\write\rfile{\noexpand\item{}#1}} %now unnecessary
\def\startrefs#1{\immediate\openout\rfile=refs.tmp\refno=#1}
\def\xref{\expandafter\xr@f}\def\xr@f[#1]{#1}
\def\refs#1{\count255=1[\r@fs #1{\hbox{}}]}
\def\r@fs#1{\ifx\und@fined#1\message{reflabel \string#1 is undefined.}%
\nref#1{need to supply reference \string#1.}\fi%
\vphantom{\hphantom{#1}}\edef\next{#1}\ifx\next\em@rk\def\next{}%
\else\ifx\next#1\ifodd\count255\relax\xref#1\count255=0\fi%
\else#1\count255=1\fi\let\next=\r@fs\fi\next}
%

%
% this is ugly, but moore insists
\newwrite\ffile\global\newcount\figno \global\figno=1
\def\fig{fig.~\the\figno\nfig}
\def\nfig#1{\xdef#1{fig.~\the\figno}%
\writedef{#1\leftbracket fig.\noexpand~\the\figno}%
\ifnum\figno=1\immediate\openout\ffile=figs.tmp\fi\chardef\wfile=\ffile%
\immediate\write\ffile{\noexpand\medskip\noexpand\item{Fig.\ \the\figno. }
\reflabeL{#1\hskip.55in}\pctsign}\global\advance\figno by1\findarg}
\def\xfig{\expandafter\xf@g}\def\xf@g fig.\penalty\@M\ {}
\def\figs#1{figs.~\f@gs #1{\hbox{}}}
\def\f@gs#1{\edef\next{#1}\ifx\next\em@rk\def\next{}\else
\ifx\next#1\xfig #1\else#1\fi\let\next=\f@gs\fi\next}
\newwrite\lfile
{\escapechar-1\xdef\pctsign{\string\%}\xdef\leftbracket{\string\{}
\xdef\rightbracket{\string\}}\xdef\numbersign{\string\#}}

\def\writestop{\def\writestoppt{\immediate\write\lfile{\string\pageno%
\the\pageno\string\startrefs\leftbracket\the\refno\rightbracket%
\string\def\string\secsym\leftbracket\secsym\rightbracket%
\string\secno\the\secno\string\meqno\the\meqno}\immediate\closeout\lfile}}
\def\writestoppt{}\def\writedef#1{}
\def\seclab#1{\xdef #1{\the\secno}\writedef{#1\leftbracket#1}\wrlabeL{#1=#1}}
\def\subseclab#1{\xdef #1{\secsym\the\subsecno}%
\writedef{#1\leftbracket#1}\wrlabeL{#1=#1}}
\newwrite\tfile \def\writetoca#1{}
\def\leaderfill{\leaders\hbox to 1em{\hss.\hss}\hfill}
%        use this to write file with table of contents
\def\writetoc{\immediate\openout\tfile=toc.tmp
   \def\writetoca##1{{\edef\next{\write\tfile{\noindent ##1
   \string\leaderfill {\noexpand\number\pageno} \par}}\next}}}
%       and this lists table of contents on second pass

%
\catcode`\@=12 % at signs are no longer letters
%
%        Unpleasantness in calling in abstract and title fonts
\edef\tfontsize{\ifx\answ\bigans scaled\magstep3\else scaled\magstep4\fi}
\font\titlerm=cmr10 \tfontsize \font\titlerms=cmr7 \tfontsize
\font\titlermss=cmr5 \tfontsize \font\titlei=cmmi10 \tfontsize
\font\titleis=cmmi7 \tfontsize \font\titleiss=cmmi5 \tfontsize
\font\titlesy=cmsy10 \tfontsize \font\titlesys=cmsy7 \tfontsize
\font\titlesyss=cmsy5 \tfontsize \font\titleit=cmti10 \tfontsize
\skewchar\titlei='177 \skewchar\titleis='177 \skewchar\titleiss='177
\skewchar\titlesy='60 \skewchar\titlesys='60 \skewchar\titlesyss='60
\def\titlefont{\def\rm{\fam0\titlerm}% switch to title font
\textfont0=\titlerm \scriptfont0=\titlerms \scriptscriptfont0=\titlermss
\textfont1=\titlei \scriptfont1=\titleis \scriptscriptfont1=\titleiss
\textfont2=\titlesy \scriptfont2=\titlesys \scriptscriptfont2=\titlesyss
\textfont\itfam=\titleit \def\it{\fam\itfam\titleit}\rm}
 \ifx\answ\bigans\else scaled\magstep1\fi
\ifx\answ\bigans\def\abstractfont{\tenpoint}\else
\font\abssl=cmsl10 scaled \magstep1
\font\absrm=cmr10 scaled\magstep1 \font\absrms=cmr7 scaled\magstep1
\font\absrmss=cmr5 scaled\magstep1 \font\absi=cmmi10 scaled\magstep1
\font\absis=cmmi7 scaled\magstep1 \font\absiss=cmmi5 scaled\magstep1
\font\abssy=cmsy10 scaled\magstep1 \font\abssys=cmsy7 scaled\magstep1
\font\abssyss=cmsy5 scaled\magstep1 \font\absbf=cmbx10 scaled\magstep1
\skewchar\absi='177 \skewchar\absis='177 \skewchar\absiss='177
\skewchar\abssy='60 \skewchar\abssys='60 \skewchar\abssyss='60
\def\abstractfont{\def\rm{\fam0\absrm}% switch to abstract font
\textfont0=\absrm \scriptfont0=\absrms \scriptscriptfont0=\absrmss
\textfont1=\absi \scriptfont1=\absis \scriptscriptfont1=\absiss
\textfont2=\abssy \scriptfont2=\abssys \scriptscriptfont2=\abssyss
\textfont\itfam=\bigit \def\it{\fam\itfam\bigit}\def\footnotefont{\tenpoint}%
\textfont\slfam=\abssl \def\sl{\fam\slfam\abssl}%
\textfont\bffam=\absbf \def\bf{\fam\bffam\absbf}\rm}\fi
\def\tenpoint{\def\rm{\fam0\tenrm}% switch back to 10-point type
\textfont0=\tenrm \scriptfont0=\sevenrm \scriptscriptfont0=\fiverm
\textfont1=\teni  \scriptfont1=\seveni  \scriptscriptfont1=\fivei
\textfont2=\tensy \scriptfont2=\sevensy \scriptscriptfont2=\fivesy
\textfont\itfam=\tenit
\def\it{\fam\itfam\tenit}\def\footnotefont{\ninepoint}%
\textfont\bffam=\tenbf \def\bf{\fam\bffam\tenbf}\def\sl{\fam\slfam\tensl}\rm}
\font\ninerm=cmr9 \font\sixrm=cmr6 \font\ninei=cmmi9 \font\sixi=cmmi6
\font\ninesy=cmsy9 \font\sixsy=cmsy6 \font\ninebf=cmbx9
\font\nineit=cmti9 \font\ninesl=cmsl9 \skewchar\ninei='177
\skewchar\sixi='177 \skewchar\ninesy='60 \skewchar\sixsy='60
\def\ninepoint{\def\rm{\fam0\ninerm}% switch to footnote font
\textfont0=\ninerm \scriptfont0=\sixrm \scriptscriptfont0=\fiverm
\textfont1=\ninei \scriptfont1=\sixi \scriptscriptfont1=\fivei
\textfont2=\ninesy \scriptfont2=\sixsy \scriptscriptfont2=\fivesy
\textfont\itfam=\ninei \def\it{\fam\itfam\nineit}\def\sl{\fam\slfam\ninesl}%
\textfont\bffam=\ninebf \def\bf{\fam\bffam\ninebf}\rm}
%
%---------------------------------------------------------------------
%

\hyphenation{anom-aly anom-alies coun-ter-term coun-ter-terms}
\def\inv{^{\raise.15ex\hbox{${\scriptscriptstyle -}$}\kern-.05em 1}}

\def\Dsl{\,\raise.15ex\hbox{/}\mkern-13.5mu D} %this one can be subscripted
\def\dsl{\raise.15ex\hbox{/}\kern-.57em\partial}

\font\bigit=cmti10 scaled \magstep1
 %pound sterling
\def\lspace{\ifx\answ\bigans{}\else\qquad\fi}
\def\lbspace{\ifx\answ\bigans{}\else\hskip-.2in\fi} % $$\lbspace...$$
\def\boxeqn#1{\vcenter{\vbox{\hrule\hbox{\vrule\kern3pt\vbox{\kern3pt
           \hbox{${\displaystyle #1}$}\kern3pt}\kern3pt\vrule}\hrule}}}
\def\mbox#1#2{\vcenter{\hrule \hbox{\vrule height#2in
               \kern#1in \vrule} \hrule}}  %e.g. \mbox{.1}{.1}
%       matters of taste
%\def\tilde{\widetilde} \def\bar{\overline} \def\hat{\widehat}
%
% some sample definitions
  %     curly letters

\def\e#1{{\rm e}^{^{\textstyle#1}}}

\def\darr#1{\raise1.5ex\hbox{$\leftrightarrow$}\mkern-16.5mu #1}
\def\lie{\hbox{\it\$}} %pound sterling

\def\half{{\textstyle{1\over2}}} %puts a small half in a displayed eqn
\def\roughly#1{\raise.3ex\hbox{$#1$\kern-.75em\lower1ex\hbox{$\sim$}}}

%\input harvmac.tex

%%temporary additional macros
% \input macros.tex
% April 16 -- NN

%%%%%%%%%%%%%%%%%%%%%  Rublenye bukvy   %%%%%%%%%%%%%%%%%%%%%%%%
\def\IB{\relax\hbox{$\inbar\kern-.3em{\rm B}$}}
\def\IC{\relax\hbox{$\inbar\kern-.3em{\rm C}$}}
\def\ID{\relax\hbox{$\inbar\kern-.3em{\rm D}$}}
\def\IE{\relax\hbox{$\inbar\kern-.3em{\rm E}$}}
\def\IF{\relax\hbox{$\inbar\kern-.3em{\rm F}$}}
\def\IG{\relax\hbox{$\inbar\kern-.3em{\rm G}$}}
\def\IGa{\relax\hbox{${\rm I}\kern-.18em\Gamma$}}
\def\IH{\relax{\rm I\kern-.18em H}}
\def\IK{\relax{\rm I\kern-.18em K}}
\def\II{\relax{\rm I\kern-.18em I}}
\def\IL{\relax{\rm I\kern-.18em L}}
\def\IP{\relax{\rm I\kern-.18em P}}
\def\IR{\relax{\rm I\kern-.18em R}}
\def\IZ{\relax\ifmmode\mathchoice {\hbox{\cmss Z\kern-.4em Z}}{\hbox{\cmss
Z\kern-.4em Z}} {\lower.9pt\hbox{\cmsss Z\kern-.4em Z}}
{\lower1.2pt\hbox{\cmsss Z\kern-.4em Z}}\else{\cmss Z\kern-.4em Z}\fi}

\def\IB{\relax{\rm I\kern-.18em B}}
\def\IC{{\relax\hbox{$\inbar\kern-.3em{\rm C}$}}}
\def\ID{\relax{\rm I\kern-.18em D}}
\def\IE{\relax{\rm I\kern-.18em E}}
\def\IF{\relax{\rm I\kern-.18em F}}

%%%%%%%%%%%%%%%%%%%% Calligraphic letters  %%%%%%%%%%%%%%%%%%%%%%%

%%%%%%%%%%%%%%%%%%%%%%%%%% Derivatives  %%%%%%%%%%%%%%%%%%%%%%%%
\def\p{\partial}

%%Beltrami

%%%%%%%%%%%%%%%%%%%% letters with bar %%%%%%%%%%%%%%%%%%%%%%%%%%

\def\z{{\bar {z}}}

%%%%%%%%%%%%%%%%%%%%%%%%%%% Math symbols %%%%%%%%%%%%%%%%%%%%%%%

\def\s{\lies}

%%%%%%%%%%%%%%%%%%%%% Short Cuts %%%%%%%%%%%%%%%%%%%%%%%

\def\half {{1\over 2}}

%%%%%%%%%%%%%%%%%% Greek %%%%%%%%%%%%%%%%%%%%%%

\def\a{\alpha}
\def\b{\beta}
\def\g{\gamma}  \def\G{\Gamma}
\def\d{\delta}  
\def\m{\mu}

\def\l{\lambda} 
\def\k{\kappa}
\def\e{\epsilon}

%%%%%%%%%%%%%%%%%% Big ( )  %%%%%%%%%%%%%%%%%%%%%%
\def\|{\Big|}
\def\({\Big(}   \def\){\Big)}
\def\[{\Big[}   \def\]{\Big]}

%%%%%%%%%%%%%%%%%% Text %%%%%%%%%%%%%%%%%%%%%%

%%%%%%%%%%%%% References %%%%%%%%%%%%%%%%%%%%

% refs with #1=authors, #2=title, #3=publ.ref, #4=hep no :
%\lref\NAME{\paper
%{Authors}{Title(in \it)}{\PLB{No.}{Year}{page},}
%{\hh 0006036 (in\tt)}.}

%\def\hh#1{hep-th/{\it #1}}

% journal~{\bf no.} (year) page

%%%%%%%%%%%%%%%%%%% Something to deal with sub-sub-sections
%%%%%%%%%%%%%%%%%%%%%%%%%%%%%%%%%%%%%%%%%%%%%%%

\def\unlockat{\catcode`\@=11}
\def\lockat{\catcode`\@=12}

\unlockat

% Something to deal with sub-sub-sections

\def\newsec#1{\global\advance\secno by1\message{(\the\secno. #1)}
\global\subsecno=0\global\subsubsecno=0\eqnres@t\noindent {\bf\the\secno. #1}
\writetoca{{\secsym} {#1}}\par\nobreak\medskip\nobreak}
\global\newcount\subsecno \global\subsecno=0
\def\subsec#1{\global\advance\subsecno by1\message{(\secsym\the\subsecno.
#1)}
\ifnum\lastpenalty>9000\else\bigbreak\fi\global\subsubsecno=0
\noindent{\it\secsym\the\subsecno. #1}
\writetoca{\string\quad {\secsym\the\subsecno.} {#1}}
\par\nobreak\medskip\nobreak}
\global\newcount\subsubsecno \global\subsubsecno=0
\def\subsubsec#1{\global\advance\subsubsecno by1
\message{(\secsym\the\subsecno.\the\subsubsecno. #1)}
\ifnum\lastpenalty>9000\else\bigbreak\fi
\noindent\quad{\secsym\the\subsecno.\the\subsubsecno.}{#1}
\writetoca{\string\qquad{\secsym\the\subsecno.\the\subsubsecno.}{#1}}
\par\nobreak\medskip\nobreak}

\def\subsubseclab#1{\DefWarn#1\xdef #1{\noexpand\hyperref{}{subsubsection}%
{\secsym\the\subsecno.\the\subsubsecno}%
{\secsym\the\subsecno.\the\subsubsecno}}%
\writedef{#1\leftbracket#1}\wrlabeL{#1=#1}}% Macros for boxes
\lockat

%why???\font\manual=manfnt
\def\dbend{\lower3.5pt\hbox{\manual\char127}}

%%%%%%%%%%%%%%%%%%% Macros for boxes %%%%%%%%%%%%%%%%%%

\def\boxit#1{\vbox{\hrule\hbox{\vrule\kern8pt
\vbox{\hbox{\kern8pt}\hbox{\vbox{#1}}\hbox{\kern8pt}}
\kern8pt\vrule}\hrule}}

\def\mathboxit#1{\vbox{\hrule\hbox{\vrule\kern8pt\vbox{\kern8pt
\hbox{$\displaystyle #1$}\kern8pt}\kern8pt\vrule}\hrule}}

\overfullrule=0pt

%%%%%%%%%%%%%%%%%%%%%%%%%% Derivatives  %%%%%%%%%%%%%%%%%%%%%%%%

\def\p{\partial}

%%Beltrami

%%%%%%%%%%%%%%%%%%%% letters with bar %%%%%%%%%%%%%%%%%%%%%%%%%%

\def\z{{\bar {z}}}

%%%%%%%%%%%%%%%%%%%%% Short Cuts %%%%%%%%%%%%%%%%%%%%%%%

\def\half {{1\over 2}}

%%%%%%%%%%%%%%%%%% Greek %%%%%%%%%%%%%%%%%%%%%%

\def\a{\alpha}
\def\b{\beta}
\def\g{\gamma}  \def\G{\Gamma}
\def\d{\delta}  
\def\m{\mu}

\def\l{\lambda} 
\def\k{\kappa}
\def\e{\epsilon}

\def\a{\alpha}
\def\b{\beta}
\def\d{\delta}

\def\m{\mu}

\def\s{\sigma}
\def\l{\lambda}

\def\k{\kappa}

\def\t{\theta}

%%%%%%%%%%%%%%%%%% Big ( )  %%%%%%%%%%%%%%%%%%%%%%

\def\|{\Big|}
\def\({\Big(}   \def\){\Big)}
\def\[{\Big[}   \def\]{\Big]}

%%%%%%%%%%%%%%%%%% Text %%%%%%%%%%%%%%%%%%%%%%

%%%%%%%%%%%%% References %%%%%%%%%%%%%%%%%%%%

% refs with #1=authors, #2=title, #3=publ.ref, #4=hep no :
%\lref\NAME{\paper
%{Authors}{Title(in \it)}{\PLB{No.}{Year}{page},}
%{\hh 0006036 (in\tt)}.}

%\def\hh#1{hep-th/{\it #1}}

% journal~{\bf no.} (year) page

%%%%%%%%%%%%%%%%%%%%%%%%%%%%%%%%%%%%%%%%%%%%%%%%%%%%%%%%%%%%%%%%

\Title{\vbox{\hbox{YITP-SB-01-71}\hbox{NYU-TH/01/12/02}}} 
{\vbox{ 
\centerline{Covariant Quantization of Superstrings}
\vskip .2cm
\centerline{Without Pure Spinor Constraints}}} 
\medskip\centerline{P.A. Grassi$^{~a,}$\foot{pgrassi@insti.physics.sunysb.edu}, 
G. Policastro$^{~b,}$\foot{policast@cibslogin.sns.it}, 
M. Porrati$^{~c,}$\foot{mp9@scires.nyu.edu}
and 
P. van Nieuwenhuizen$^{~a,}$\foot{vannieu@insti.physics.sunysb.edu}}
%M. Porrati\foot{e-mail: {pag5, gp27, mp9}@nyu.edu}} 
\medskip 
\centerline{$^{(a)}$ {\it C.N. Yang Institute for Theoretical Physics,} }
\centerline{\it State University of New York at Stony Brook, 
NY 11794-3840, USA}
\vskip .3cm
\centerline{$^{(b)}$ {\it Scuola Normale Superiore,} }
\centerline{\it Piazza dei Cavalieri 7, Pisa, 56126, Italy}
\vskip .3cm
\centerline{$^{(c)}$ {\it New York University Physics Department,} }
\centerline{\it 4 Washington Place, New York, 10003, NY, USA}
\medskip
\vskip  .5cm
\noindent
We construct a covariant quantum superstring, extending 
Berkovits' approach by introdu\-cing new ghosts to relax the pure spinor constraints. 
The central charge of the underlying Kac-Moody algebra, which would lead to an 
anomaly in the BRST charge, is treated as a new generator with a new $b-c$ system.   
We construct a nilpotent BRST current, an anomalous ghost current and an anomaly-free 
energy-momentum tensor. For open superstrings, we find the correct massless spectrum.
In addition, we construct a Lorentz invariant $B$-field to be used for the computation of the 
integrated vertex operators and amplitudes. 

\Date{\ December 2001}

% References

\lref\lrp{
U.~Lindstr\"om, M.~Ro\v cek, and P.~van Nieuwenhuizen, in preparation. 
}

\lref\pr{
P. van Nieuwenhuizen, in {\it Supergravity `81}, 
Proceedings First School on Supergravity, Cambridge University Press, 1982, page 165.
}

\lref\polc{
J.~Polchinski,
{\it String Theory. Vol. 1: An Introduction To The Bosonic String,}
{\it String Theory. Vol. 2: Superstring Theory And Beyond,}
{\it  Cambridge, UK: Univ. Pr. (1998) 531 p}.
}

\lref\superstring{
M.~B.~Green and  J.~H.~Schwarz, {\it Covariant Description Of Superstrings,} 
Phys.\  Lett.\ {\bf B136} (1984) 367; M.~B.~Green and J.~H.~Schwarz,
  {\it Properties Of The Covariant Formulation Of Superstring Theories,}
  Nucl.\ Phys.\ {\bf B243} (1984) 285\semi
M.~B.~Green and C.~M.~Hull, QMC/PH/89-7
{\it Presented at Texas A and M Mtg. on String Theory, College
  Station, TX, Mar 13-18, 1989}\semi
R.~Kallosh and M.~Rakhmanov, Phys.\ Lett.\  {\bf B209} (1988) 233\semi
%M.~B.~Green and C.~M.~Hull, ``The Brst Cohomology Of An N=1 Superparticle,''
%{\it  In *College Station 1990, Proceedings, Strings 90* 133-147. }; 
U. ~Lindstr\"om, M.~Ro\v cek, W.~Siegel, 
P.~van Nieuwenhuizen and A.~E.~van de Ven, Phys. Lett. {\bf B224} (1989) 
285, Phys. Lett. {\bf B227}(1989) 87, and Phys. Lett. {\bf B228}(1989) 53; 
S.~J.~Gates, M.~T.~Grisaru, U.~Lindstr\"om, M.~Ro\v cek, W.~Siegel, 
P.~van Nieuwenhuizen and A.~E.~van de Ven,
{\it Lorentz Covariant Quantization Of The Heterotic Superstring,}
Phys.\ Lett.\  {\bf B225} (1989) 44; 
A.~Mikovic, M.~Rocek, W.~Siegel, P.~van Nieuwenhuizen, J.~Yamron and
A.~E.~van de Ven, Phys.\ Lett.\  {\bf B235} (1990) 106; 
U.~Lindstr\"om, M.~Ro\v cek, W.~Siegel, P.~van Nieuwenhuizen and
A.~E.~van de Ven, 
{\it Construction Of The Covariantly Quantized Heterotic Superstring,}
Nucl.\ Phys.\  {\bf B330} (1990) 19 \semi
F. Bastianelli, G. W. Delius and E. Laenen, Phys. \ Lett. \ {\bf
  B229}, 223 (1989)\semi
R.~Kallosh, Nucl.\ Phys.\ Proc.\ Suppl.\  {\bf 18B}
  (1990) 180 \semi
M.~B.~Green and C.~M.~Hull, Mod.\ Phys.\ Lett.\  {\bf A5} (1990) 1399\semi 
M.~B.~Green and C.~M.~Hull, Nucl.\ Phys.\  {\bf B344} (1990) 115\semi
F.~Essler, E.~Laenen, W.~Siegel and J.~P.~Yamron, Phys.\ Lett.\  {\bf B254} (1991) 411\semi 
  F.~Essler, M.~Hatsuda, E.~Laenen, W.~Siegel, J.~P.~Yamron, T.~Kimura
  and A.~Mikovic, 
  Nucl.\ Phys.\  {\bf B364} (1991) 67\semi 
J.~L.~Vazquez-Bello,
  Int.\ J.\ Mod.\ Phys.\  {\bf A7} (1992) 4583\semi
E. Bergshoeff, R. Kallosh and A. Van Proeyen, ``Superparticle
  actions and gauge fixings'', Class.\ Quant.\ Grav {\bf 9} 
  (1992) 321\semi
C.~M.~Hull and J.~Vazquez-Bello, Nucl.\ Phys.\  {\bf B416}, (1994) 173 [hep-th/9308022]\semi
P.~A.~Grassi, G.~Policastro and M.~Porrati,
{\it Covariant quantization of the Brink-Schwarz superparticle,}
Nucl.\ Phys.\ B {\bf 606}, 380 (2001)
[arXiv:hep-th/0009239].
}

\lref\bv{
N. Berkovits and C. Vafa,
{\it $N=4$ Topological Strings}, Nucl. Phys. B433 (1995) 123, 
hep-th/9407190.}

\lref\fourreview{N. Berkovits,  {\it Covariant Quantization Of
The Green-Schwarz Superstring In A Calabi-Yau Background,}
Nucl. Phys. {\bf B431} (1994) 258, ``A New Description Of The Superstring,''
Jorge Swieca Summer School 1995, p. 490, hep-th/9604123.}

%\OoguriPS
\lref\OoguriPS{
H.~Ooguri, J.~Rahmfeld, H.~Robins and J.~Tannenhauser,
{\it Holography in superspace,}
JHEP {\bf 0007}, 045 (2000)
[arXiv:hep-th/0007104].
%%CITATION = HEP-TH 0007104;%%
}

\lref\bvw{
N.~Berkovits, C.~Vafa and E.~Witten,
{\it Conformal field theory of AdS background with Ramond-Ramond flux,}
JHEP {\bf 9903}, 018 (1999)
[arXiv:hep-th/9902098].
%%CITATION = HEP-TH 9902098;%%
}
\lref\wittwi{
E.~Witten,
{\it An Interpretation Of Classical Yang-Mills Theory,}
Phys.\ Lett.\ B {\bf 77}, 394 (1978); 
E.~Witten,
{\it Twistor - Like Transform In Ten-Dimensions,}
Nucl.\ Phys.\ B {\bf 266}, 245 (1986)}
%J.~P.~Harnad and S.~Shnider,
%{\it Constraints And Field Equations For Ten-Dimensional 
%Super-Yang-Mills Theory,}
%Commun.\ Math.\ Phys.\  {\bf 106}, 183 (1986)\semi

\lref\SYM{
W.~Siegel,
{\it Superfields In Higher Dimensional Space-Time,}
Phys.\ Lett.\ B {\bf 80}, 220 (1979)\semi
B.~E.~Nilsson,
{\it Pure Spinors As Auxiliary Fields In The Ten-Dimensional 
Supersymmetric Yang-Mills Theory,}
Class.\ Quant.\ Grav.\  {\bf 3}, L41 (1986); 
B.~E.~Nilsson,
{\it Off-Shell Fields For The Ten-Dimensional Supersymmetric 
Yang-Mills Theory,} GOTEBORG-81-6\semi
S.~J.~Gates and S.~Vashakidze,
{\it On D = 10, N=1 Supersymmetry, Superspace Geometry And Superstring Effects,}
Nucl.\ Phys.\ B {\bf 291}, 172 (1987)\semi
M.~Cederwall, B.~E.~Nilsson and D.~Tsimpis,
{\it The structure of maximally supersymmetric Yang-Mills theory:  
Constraining higher-order corrections,}
JHEP {\bf 0106}, 034 (2001)
[arXiv:hep-th/0102009]; 
M.~Cederwall, B.~E.~Nilsson and D.~Tsimpis,
{\it D = 10 superYang-Mills at O(alpha**2),}
JHEP {\bf 0107}, 042 (2001)
[arXiv:hep-th/0104236].
}
\lref\har{
J.~P.~Harnad and S.~Shnider,
{\it Constraints And Field Equations For Ten-Dimensional 
Super-Yang-Mills Theory,}
Commun.\ Math.\ Phys.\  {\bf 106}, 183 (1986).
}
\lref\wie{
%\lref\WiegmannHN{
P.~B.~Wiegmann,
{\it Multivalued Functionals And Geometrical Approach 
For Quantization Of Relativistic Particles And Strings,} 
Nucl.\ Phys.\ B {\bf 323}, 311 (1989).
%%CITATION = NUPHA,B323,311;%%
}
\lref\purespinors{\'E. Cartan, {\it Lecons sur la th\'eorie des spineurs}, 
Hermann, Paris (1937)\semi
C. Chevalley, {\it The algebraic theory of Spinors}, 
Columbia Univ. Press., New York\semi
 R. Penrose and W. Rindler, 
{\it Spinors and Space-Time}, Cambridge Univ. Press, Cambridge (1984) 
\semi
P. Budinich and A. Trautman, {\it The spinorial chessboard}, Springer, 
New York (1989).
}
\lref\coset{
P.~Furlan and R.~Raczka,
{\it Nonlinear Spinor Representations,}
J.\ Math.\ Phys.\  {\bf 26}, 3021 (1985)\semi
%%CITATION = JMAPA,26,3021;%%
A.~S.~Galperin, P.~S.~Howe and K.~S.~Stelle,
{\it The Superparticle and the Lorentz group,}
Nucl.\ Phys.\ B {\bf 368}, 248 (1992)
[arXiv:hep-th/9201020].
%%CITATION = HEP-TH 9201020;%%
}

%
%\lref\superspace{
%S.~J.~Gates, M.~T.~Grisaru, M.~Rocek and W.~Siegel,
%{\it Superspace, Or One Thousand 
%And One Lessons In Supersymmetry,''}
%Front.\ Phys.\  {\bf 58}, 1 (1983)
%[arXiv:hep-th/0108200].}
%
\lref\GS{M.B. Green, J.H. Schwarz, and E. Witten, {\it Superstring Theory,} 
 vol. 1, chapter 5 (Cambridge U. Press, 1987).  
}
\lref\carlip{S. Carlip, 
{\it Heterotic String Path Integrals with the Green-Schwarz 
Covariant Action}, Nucl. Phys. B284 (1987) 365 \semi R. Kallosh, 
{\it Quantization of Green-Schwarz Superstring}, Phys. Lett. B195 (1987) 369.} 
 \lref\john{G. Gilbert and 
D. Johnston, {\it Equivalence of the Kallosh and Carlip Quantizations 
of the Green-Schwarz Action for the Heterotic String}, Phys. Lett. B205 
(1988) 273.} 
\lref\csm{W. Siegel, {\it Classical Superstring Mechanics}, Nucl. Phys. B263 (1986) 
93\semi 
W.~Siegel, {\it Randomizing the Superstring}, Phys. Rev. D 50 (1994), 2799.
}   
\lref\sok{E. Sokatchev, {\it 
Harmonic Superparticle}, Class. Quant. Grav. 4 (1987) 237\semi 
E.R. Nissimov and S.J. Pacheva, {\it Manifestly Super-Poincar\'e 
Covariant Quantization of the Green-Schwarz Superstring}, 
Phys. Lett. B202 (1988) 325\semi 
R. Kallosh and M. Rakhmanov, {\it Covariant Quantization of the 
Green-Schwarz Superstring}, Phys. Lett. B209 (1988) 233.}  
\lref\many{S.J. Gates Jr, M.T. Grisaru, 
U. Lindstrom, M. Rocek, W. Siegel, P. van Nieuwenhuizen and 
A.E. van de Ven, {\it Lorentz-Covariant Quantization of the Heterotic 
Superstring}, Phys. Lett. B225 (1989) 44\semi 
R.E. Kallosh, {\it Covariant Quantization of Type IIA,B 
Green-Schwarz Superstring}, Phys. Lett. B225 (1989) 49\semi 
M.B. Green and C.M. Hull, {\it Covariant Quantum Mechanics of the 
Superstring}, Phys. Lett. B225 (1989) 57.}  
 \lref\fms{D. Friedan, E. Martinec and S. Shenker, 
{\it Conformal Invariance, Supersymmetry and String Theory}, 
Nucl. Phys. B271 (1986) 93.}
\lref\kawai{
T.~Kawai,
{\it Remarks On A Class Of BRST Operators,}
Phys.\ Lett.\ B {\bf 168}, 355 (1986).}
 \lref\ufive{N. Berkovits, {\it 
Quantization of the Superstring with Manifest U(5) Super-Poincar\'e 
Invariance}, Phys. Lett. B457 (1999) 94, hep-th/9902099.}  
\lref\BerkovitsRB{ N.~Berkovits, 
{\it Covariant quantization of the superparticle 
using pure spinors,} [hep-th/0105050].  
%%CITATION =HEP-TH 0105050;%% 
} 

%%% berkovits %%%%

%\BerkovitsFE
\lref\BerkovitsFE{
N.~Berkovits,
{\it Super-Poincar\'e covariant quantization of the superstring,}
JHEP { 0004}, 018 (2000)
[hep-th/0001035].
}

%\BerkovitsPH
\lref\BerkovitsPH{
N.~Berkovits and B.~C.~Vallilo,
{\it Consistency of super-Poincar\'e covariant superstring tree amplitudes,}
JHEP { 0007}, 015 (2000)
[hep-th/0004171].
}

%\BerkovitsNN
\lref\BerkovitsNN{
N.~Berkovits,
{\it Cohomology in the pure spinor formalism for the superstring,}
JHEP { 0009}, 046 (2000)
[hep-th/0006003].
}

%\BerkovitsWM
\lref\BerkovitsWM{
N.~Berkovits,
{\it Covariant quantization of the superstring,}
Int.\ J.\ Mod.\ Phys.\ A { 16}, 801 (2001)
[hep-th/0008145].
}

%\BerkovitsYR
\lref\BerkovitsYR{
N.~Berkovits and O.~Chandia,
{\it Superstring vertex operators in an AdS(5) x S(5) background,}
Nucl.\ Phys.\ B {\bf 596}, 185 (2001)
[hep-th/0009168].
}
%\BerkovitsZY
\lref\BerkovitsZY{
N.~Berkovits,
{\it The Ten-dimensional Green-Schwarz 
superstring is a twisted Neveu-Schwarz-Ramond string,}
Nucl.\ Phys.\ B {\bf 420}, 332 (1994)
[arXiv:hep-th/9308129].
%%CITATION = HEP-TH 9308129;%%
}

%\BerkovitsUS
\lref\BerkovitsUS{
N.~Berkovits,
{\it Relating the RNS and pure spinor formalisms for the superstring,}
hep-th/0104247.
}

%\BerkovitsMX
\lref\BerkovitsMX{
N.~Berkovits and O.~Chandia,
{\it Lorentz invariance of the pure spinor BRST cohomology 
for the  superstring,}
hep-th/0105149.
}

%\WittenZZ
\lref\WittenZZ{
E.~Witten,
{\it Mirror manifolds and topological field theory,}
hep-th/9112056.
%%CITATION = HEP-TH 9112056;%%
}

\lref\wichen{
E.~Witten,
{\it Chern-Simons gauge theory as a string theory,}
arXiv:hep-th/9207094.
%%CITATION = HEP-TH 9207094;%%
}

%\lref\bgi{
%C.~Becchi, S.~Giusto and C.~Imbimbo,
%{\it The holomorphic anomaly of topological strings,}
%Fortsch.\ Phys.\  {\bf 47}, 195 (1999)
%[hep-th/9801100]\semi
%C.~Becchi, S.~Giusto and C.~Imbimbo, 
%{\it Topological B models}, unpublished. 
%}

%\howe
\lref\howe{P.S. Howe, {\it Pure Spinor Lines in Superspace and 
Ten-Dimensional Supersymmetric Theories}, 
Phys. Lett. B258 (1991) 141, Addendum-ibid.B259 (1991) 511\semi 
P.S. Howe, {\it Pure Spinors, Function Superspaces and Supergravity 
Theories in Ten Dimensions and Eleven Dimensions}, Phys. Lett. B273 (1991) 
90.}

%%%%%%%%%%%%%%%%%%%%%%%%%%%%%%%%%%%%%%%%%%%%%%%%%%%%%%%%%%%%%%%%%%%
\baselineskip14pt

\newsec{Introduction and Summary} 

Recently, a new formulation of superstrings was developed which is
explicitly super-Poincar\'e invariant in $9+1$ dimensions
\BerkovitsFE\ \BerkovitsPH\ \BerkovitsNN\ \BerkovitsWM.  It is based
on a {\it free} conformal field theory on the world-sheet and a
nilpotent BRST charge $Q$ which defines the physical vertices as
representatives of its cohomology. In addition to the conventional
variables $x^m$ and $\t^\a$ of the Green-Schwarz formalism, a
conjugate momentum $p_\a$ for $\t^\a$ and a set of commuting ghost fields
$\l^\a$ are introduced. The latter are complex Weyl spinors satisfying the
pure spinor conditions $\l^\a \g^m_{\a\b} \l^\b = 0$ \BerkovitsFE\
\howe. This equation can be solved by decomposing $\lambda$ with
respect to a non-compact $U(5)$ subgroup of $SO(9,1)$ into a singlet
$\underline{1}$, a vector $\underline{5}$, and a tensor
$\underline{10}$. The vector can be expressed in terms of the singlet
and tensor, hence there are 11 independent complex variables in
$\l^\a$.

The pure spinors $\l^\a$ are needed to cancel the 
central charge of the conformal algebra ($+10$ from the bosonic coordinates $x^m$, $-32$ from
the spinor variables $p_\a$ and $\t^\a$, and $+22$ from the pure
spinors $\l^\a$), to obtain the correct  
double poles in the Lorentz algebra and, last but not least, to render the operator  $Q$ nilpotent. 

 Of course, the problem of the covariant quantization of superstrings
is one of the fundamental problems in string theory. The subject
has a long history.  We give in \superstring\ some of the early papers
on the subject which we used, but the list is far from complete.

As shown by Berkovits in \BerkovitsNN , the cohomology of the BRST charge contains 
exactly the physical spectrum of the superstring and, in particular
for the massless states, it provides  
the covariant equations of motion, those of super Yang-Mills theory in $10$ dimensions 
for the open superstring or those of $N=2$ supergravity for closed superstrings. 
For massive states he could not use the formalism based on the $U(5)$-like subgroup; 
it was possible to use an $SO(8)$ subgroup, but this introduced an infinite set of 
ghost-for-ghosts although the cohomology did not depend on these fields. 

Although Berkovits's approach provides a way to by-pass 
many of the difficulties of the Green-Schwarz formalism for  a  super-Poincar\'e 
covariant description of superstrings,   
an explicit parametrization of pure spinors $\l^\a$ is needed at several points in his construction.  
Using such an explicit parametrization for pure spinors, 
he was able to define Lorentz currents which satisfy the covariant Lorentz algebra 
\BerkovitsMX. 
However, the solution of the pure spinor constraints breaks the
explicit covariance and some expressions cannot be written in a
covariant way. For example, the action is not covariant and also
intermediate steps in the computation of amplitudes are clearly
affected. In addition, a conjugate momentum for the pure spinors can
only be constructed in an explicit parametrization, resulting in OPE's
which are not manifestly covariant.  The tree level amplitudes are not
manifestly super-Poincar\'e invariant and massive vertices can only be
constructed in terms of a specific non-covariant parametrization of
the pure spinors.

The cohomology of \BerkovitsFE\ \BerkovitsPH\
\BerkovitsNN\ \BerkovitsWM\ is a constrained cohomology. We therefore decided 
to try to relax the pure spinor conditions and construct a 
new BRST operator such that its unconstrained cohomology 
coincides with the constrained cohomology of Berkovits's approach. The extension of the
BRST symmetry to unconstrained $\l^\a$ led us to 
enlarge the field space by adding more ghost fields: an
anticommuting vector $\xi^m$, a commuting spinor $\chi_\a$, an anticommuting vector 1-form 
$\omega_z^m$, and their
corresponding antighosts, and further an anticommuting $b-c_z$ system with conformal weights $0$
and $1$, respectively.  

Our final action for the left-moving sector is given by 
\eqn\ACTfinal{\eqalign{ S & = \int d^2z 
\Big( {1\over 2} \p x^m \bar\p x_m + p_\a \bar\p \t^\a +
 \beta_{z m} \bar \p \xi^m  + \b_{z \a} \bar\p \l^\a + \k^\a_z \bar\p \chi_\a  + c_z \, \bar\p b  + 
\omega^m_z  \bar\p \eta_m \Big)\,.
}}
Our final BRST current is given by 
\eqn\BRSTfinall{\eqalign{
 j^B_{z}& = 
\l^\a d_{z \a} - \xi^m \Pi_{z m} - \chi_\a \p_z \theta^\a - \xi^m \kappa^\a_z \g_{m \a\b} \l^\b 
-  {1\over 2} \l^\a \g^m_{\a\b} \l^\b \b_{z m} \cr
& + c_z - \half b \left( \xi^m \p_z \xi_m - {3\over 2} \chi_\a \p_z
\l^\a + \half \p_z \chi_\a \l^\a \right) 
- {1\over 2} \p_z \left( b\, \chi_\a \l^\a \right) \,, 
}}
and the energy-momentum tensor is 
\eqn\strfinal{\eqalign{
T_{zz} = -\half \Pi^m_z  \Pi_{m z} - d_{z \a} \p_z \t^\a - \b_{z m} \p_z \xi^m - \b_{z \a} \p_z \l^\a - 
\k^\a_z \p_z \chi_\a + \p_z b\, c_z +  \p_z \eta_m \, \omega^m_z \,.
}}
Because $ -\half \Pi^m_z  \Pi_{m z} - d_{z \a} \p_z \t^\a =  -\half
\p_z x^m  \p_{z} x_m - p_{z \a} \p_z \t^\a$  
we are dealing with a free conformal field theory. 
The ghost current we use is given by 
\eqn\ghostfinal{\eqalign
{
J^{gh}_z = - \left( 
\beta_{m z} \xi^m + \kappa_z^\a \chi_\a + \beta_{z \a} \l^\a + b\, c_z
+ \eta^m \omega_{z m} \right) \,.  
}}
There is also a composite field $B_{zz}$ which 
will be discussed in the text. 
%Both $j^B_z$ and $B_{zz}$ should be 
%related to the spin-3/2  currents of an $N=2$ superconformal current algebra of the NRS string 
%by twisting.  

Our main results are: 

{\it (1)} we keep manifest $SO(9,1)$ invariance at every step. 
No pure spinor constraints or $U(5)$ decompositions appear.  
As a consequence we can take $\l^\a$ to be real which solves one 
of the puzzles of \BerkovitsFE\ where $\l^\a$ had to be complex in order that the pure spinor 
constraint have a solution, but where the real $\t^\a$ transforms into complex $\l^\a$. 

{\it (2)} We construct a BRST operator with the properties one would
like it to have: its current is nilpotent  
and a primary field. Our BRST charge is based on a Kac-Moody algebra for spacetime symmetries, and 
it describes the geometry of a $D=(9,1)$ super-space. In the RNS approach the BRST charge 
describes the dynamics since it starts with $\oint c^z T_{zz} + \dots$. 

{\it (3)} We also construct a composite operator $B_{zz}$ which satisfies the fundamental 
relation $T_{zz} = [Q, B_{zz}]$ where $T_{zz}$ is given in \strfinal. We believe that 
$\oint B_{zz}$ plays the same role as the zero mode $b_0$ in the RNS approach. 
The integrated vertices $\oint dz  {\cal V}^{(0)}_z$ are constructed from the unintegrated vertices 
${\cal U}^{(1)}$ by $ {\cal V}^{(0)}_z = \oint dz B_{zz} {\cal U}^{(1)} +  [Q, X_{z}]$. 

{\it (4)} We require that the unintegrated vertex operators should
satisfy $[Q , {\cal U}^{(1)}] = 0$.  This selects all possible
deformations of the Kac-Moody algebra, in other words, the BRST
charge, with Kac-Moody generators replaced by Kac-Moody generators
plus connections, is still nilpotent.  To be physical observables these
deformations must be primary spin-1 fields, so ${\cal V}^{(0)}_z$ should 
have conformal spin $1$. This second constraint
brings in the dynamics.  For the massless sector of the open
superstring, the first constraint, namely $[Q , {\cal U}^{(1)}] = 0$,
implies that the deformations are described by a super
gauge-connection $A_\a, A_m$ and field strength $A^\a$ (usually denoted by $W^\a$) 
which satisfies the proper Bianchi
identities.  The second constraint implies the equations of motion for super
Yang-Mills theory (at the linearized level). The conditions $F_{\a\b} = F_{\a m} = 0$ 
for the superspace curvature of the gauge connection follow as
a consequence of the equations of motion.

The paper is organized in the following way: in section 2, we discuss
the construction of the BRST charge based on an extension of
Berkovits's approach and promote the central charge of the Kac-Moody
algebra to an operator. In section 3, we prove the BRST invariance of
the action and derive the boundary conditions for the open
superstring needed to maintain BRST invariance and
supersymmetry at the boundaries. Section 4 is devoted to the
construction of the energy-momentum tensor, the field $B_{zz}$, the
ghost current and to compute their OPE's. In section 5, we discuss
briefly the relation with the RNS approach and, finally, in section 6
we study the massless sector of open string. Section 7 contains some
comments and speculations.

\newsec{The BRST charge}
%\subsec{Extending the Berkovits algebra} 
 
On a flat worldsheet, the free left- and right-moving 
contributions to the classical covariant Green-Schwarz superstring action \GS~ can be written as 
\eqn\ac{
S=\int d^2 z \left(\half\p x^m \overline\p x_m + p_\a \overline\p \t^\a + \hat{p}_\a \p \hat\t^\a\right)
} 
 where $\p_z = \half (\p_\sigma - i \, \p_\tau)$ and $\bar\p_z = \half
 (\p_\sigma + i \, \p_\tau)$.  Furthermore,  
$p_\a$ (or its anti-holomorphic partner $\hat{p}_\a$) is related to $x^m$ and $\t^\a$ 
by the constraint $d_\a=0$ with \csm 
\eqn\dd{d_\a = p_\a - {1 \over 2}\p x_m \g^m_{\a\b} \t^\b - {1\over 8}\g^m_{\a\b} \g_{m\,\g\d} 
\t^\b\t^\g\p\t^\d.} 
Setting $d_\a$ equal to zero, \ac\ leads to an interacting superstring whose left-movers 
do not interact with its right-movers, and which therefore is not
equal to the Green-Schwarz superstring.  
The variables $x^m$ and $\t^\a$ are worldsheet  scalar fields. On the other hand, the conjugate 
momenta $p_\a, d_\a$ and $\Pi^m = \p x^m + {1 \over 2} \t^\a \g^m_{\a\b} \p\t^\b$ carry a  
world-sheet vector index, {\it i.e.} $p_{z, \a}, d_{z, \a},\Pi_{z, m}$. 
The index $z$ will be omitted in the following when there is no ambiguity.  
The symbols $\g^m_{\a\b}$ and $\g^{m\,\a\b}$ are real $16\times 16$ symmetric matrices which are 
the off-diagonal elements of the $32\times 32$ 
Dirac-matrices\foot{
One may use ten real $D=(9,1)$ Dirac-matrices $\Gamma^m = \{ I \otimes (i \tau_2), 
\sigma^\m \otimes \tau_1, \chi \otimes \tau_1\}$ 
where $m=0,\dots,9$ and $\m = 1,\dots,8$. The $\sigma^\m$ are eight real symmetric $16 \times 16$ 
off-diagonal Dirac matrices for $D= (8,0)$, while $\chi$ is the real
$16 \times 16$ diagonal chirality matrix in $D=8$ \pr. 
The chirality matrix in $D=(9,1)$ is then $I \otimes \tau_3$ and the $D=(9,1)$  
charge conjugation matrix $C$, satisfying  
$C \, \Gamma^m = - \Gamma^{m, T} C$, is numerically equal to $C= \Gamma^0$. If one uses spinors 
$\Psi^T = ( \a_L, \b_R)$ with spinor indices $\a_L^\a$ and $\b_{R,\dot{\b}}$, 
the index structure of the Dirac matrices and the charge conjugation matrix is 
$$ \Gamma^m = \left( \eqalign{& 0 ~~~~~~~ (\sigma^m)^{\a \dot{\b}} \cr 
 & (\tilde\sigma^m)_{\dot{\b} \g} ~~~ 0 }\right)\,, ~~~~~
C = \left( \eqalign{& 0 ~~~~~  c_{\a}^{~\dot{\b}} \cr 
 & c^{\dot{\b}}_{~ \g} ~~~ 0 }\right)\,, $$
 where $\sigma^m = \{I, \sigma^\mu, \chi \}$ and $\tilde\sigma^m = \{-I, \sigma^\mu, \chi \}$. The 
matrices $c_{\a}^{~\dot{\b}}$ and $ c^{\dot{\b}}_{~ \g} $ are
 numerically equal to $I_{16\times 16}$ and  
 $-I_{16\times 16}$, respectively. In the text, we use the symmetric matrices 
 $\g^{m \, \dot\a\dot\b} = c^{\dot{\a}}_{~\b} (\sigma^m)^{\b \dot{\b}} \, $ and $\g^{m}_{\a\b} = 
c_{\a}^{~\dot{\b}} (\tilde\sigma^m)_{\dot{\b} \b}$, and we omit the dots. The spinors 
$\a_L$ and $\b_R$ form inequivalent representations of $SO(9,1)$.  
We cannot raise and lower the spin indices with the charge conjugation matrix, but  
$\a^\a_L c_\a^{~\dot\b} \b_{R,\dot\b}$ is Lorentz invariant.  
}
and which satisfy 
$\g^m_{\a\b} \g^{n\,\b\g}+ 
\g^n_{\a\b} \g^{m\,\b\g}=2\eta^{mn}\d_\a^\g$ and 
$\g_{_{m\,(\a\b}} \g^{_{m}}_{_{\g)\d}}=0$. The latter relation makes Fierz rearrangements very easy. 

Furthermore, we have \csm\ 
\eqn\dope{\eqalign{
&d_\a(z) d_\b(w) \sim -{{\g^m_{\a\b}\Pi_m(w)}\over{z-w}},\quad 
d_\a(z) \Pi^m(w) \sim{{\g^m_{\a\b}\p\t^\b(w)} \over {z-w}}, \cr
&\Pi^m(z) \Pi^n(w) \sim- {1\over (z-w)^2} \eta^{mn}\,, \quad 
d_\a(z) \t^\b(w) \sim{1\over {z-w}} \delta^{~\b}_{\a}\,, 
}} 
where $\sim$ denotes the singular contribution to the OPE's. 
The operators $d_\a$ and $\Pi^m$ 
are invariant under the spacetime supersymmetry generated by
%\foot 
%{We have chosen the normalization $\{q_\a,q_\b\}= 
%\g^m_{a\b}\oint dz \p x_m$ to simplify comparison with RNS amplitudes.} 
\eqn\defsusy{\eqalign{
& Q_\e = \oint dz \, \e^\a j^{su}_{z \a} \,, \cr
&  j^{su}_{z \a} \equiv p_\a +\half \p x^m 
\g^m_{\a\b} \t^\b +{1\over {24}}\g^m_{\a\b} \t^\b \t^\g \g_{m\,\g\d}\p\t^\d\,, \cr
&  j^{su}_{z \a}(z)  j^{su}_{z \b}(w) \sim {
{\g^m_{\a\b} \over (z-w) } \left( \p x_m + {1\over 6} \t^\g \g_{m \g \d } \p\t^\d  \right)(w)} 
- {1 \over 3 (z-w)^2} (\g^m \t)_\a(z) (\g_m \t)_\b(w) \,
}} 
where $\e^\a$ is a constant Majorana-Weyl spinor.
The supersymmetry transformations of the fields are given by
\eqn\defsusyfields{\eqalign{
& [Q_\e, x^m] = - {1\over 2} \e^\a \g^m_{\a\b} \t^\b \,, ~~~~~~ [Q_\e, \t^\a] = \e^\a\,, \cr
& [Q_\e, p_{z\a}] = {1\over 2} \p_z x_m \g^m_{\a\b} \e^\b - {1\over 8} \g^m_{\a\b} \p_z \t^\b 
\Big( \e \g_m \t \Big)\,.
}}
The susy commutator vanishes on $p_\a$ in agreement with the vanishing of the anticommutator of the 
$\t$-dependent terms in the r.h.s. of \defsusy\ with $p_\a$. 
We require the BRST charge to be susy invariant and therefore
it is constructed on the basis $(\Pi^m_z, \p_z \t^\a, d_{z\a})$.   All the ghosts and antighosts 
will be susy inert.

The variables $(x^m, \t^\a)$ are the coordinates of $N=1$ superspace in $D=(9,1)$ dimensions 
and $(\Pi^m_z, \p_z \t^\a, d_{z\a})$ form a basis for susy-invariant super 1-forms. 
At the classical level, a general super 1-form can be written as 
\eqn\sspacehol{{\cal V}^{(0)}_z = \Pi^m_z A_m(x,\t) +  \p_z \t^\a A_\a(x,\t) + d_{z\a} A^\a(x,\t)\,,}
for the holomorphic sector, or 
\eqn\sspace{ \eqalign{
{\cal V}^{(0)}_{z\z} 
&  = \Pi^m_z \bar\Pi^{n}_\z \, G_{m \hat n}(x, \t, \hat\t) 
    + \p_z \t^\a   \bar\Pi^{n}_\z \, G_{\a \hat n}(x, \t, \hat\t) 
    + \p_z \t^\a \p_\z \hat\t^{\hat\a} \, G_{\a \hat \a}(x, \t, \hat\t) \cr
& +  \Pi^m_z \p_\z \hat\t^{\hat\a} \, G_{m \hat \a}(x, \t, \hat\t) 
    + d_{z\a}  \bar\Pi^{n}_\z \, G^\a_{~~\hat n}(x, \t, \hat\t) 
    + \Pi^m_z \hat{d}_{\z\hat\a} \, G_{m}^{~~ \hat \a}(x, \t, \hat\t) \cr
&    + d_{z\a} \p_\z \hat\t^{\hat\a} \, G^\a_{~~\hat \a}(x, \t, \hat\t) 
    + \p_z \t^\a \hat{d}_{\z\hat\a} \, G_\a^{~~\hat \a}(x, \t, \hat\t)
    +  d_{z\a} \hat{d}_{\z\hat\a} \, G^{\a \hat \a}(x, \t, \hat\t) \,.
}}
for the closed string sector. These expressions will be useful for
 the cohomology of the string. 
For the open string the functions $A_m, A_\a$ and $A^\a$ are arbitrary 
superfields of the holomorphic 
supercoordinates $(x_L^m(z), \t^\a(z))$ and  for Type II B strings the 
$G_{m \hat n}, \dots, G^{\a \hat \a}$ are superfields 
on the superspace defined by  
$x^m(z,\z)$, $ \t^\a(z)$, $\t^{\hat\a}(\z)$. The fields $x^m$ and $\t^\a$ are on-shell, so that 
we can use conformal field theory techniques. For Type II A strings, we should use 
$\hat\t_{ \a}$ instead of 
$\hat\t^{ \a}$. 
%In this way, $\Phi_z$ and 
%$\Phi_{z\z} $ are superfields and supersymmetry is kept manifest. 

In addition to the usual ten-dimensional superspace variables $x^m$ and $\t^\a$,  
Berkovits \BerkovitsFE~(see also \BerkovitsWM) 
introduced a commuting space-time spinor (worldsheet scalar) 
$\lambda^\a(z)$ satisfying the pure spinor condition \purespinors\  
\eqn\pure{\l^\a \gamma^m_{\a\b} \l^\b=0} where $m$ runs from $0$ to $9$.  
In this approach, the 16 fields $\lambda^\a$ must be complex to allow a solution of \pure. 
%and parametrize the coset $SO(9,1)/ [ SU(4) \otimes K] $ where $K$ are the conformal boosts 
%of the $D=8$ conformal group $SO(9,1)$ \coset. 
One may solve \pure\  by decomposing 
the spinor $\l^\a$ with respect to a noncompact 
subgroup of $SO(9,1)$. Introducing 5 creation  operators 
$\G^a = (\G^1 + i\, \G^2) /{2}, \dots, (\G^9 + \G^0) /{2}$ and 5 annihilation 
operators $\G_a = (\G^1 - i\, \G^2) / {2}, \dots, (\G^9 - \G^0) / {2}$, 
and defining $\l^\a$ to consist of the terms with 
an even number of these creation operators one obtains 
$\l^\a = \l_+ |0> + \half \G^{ab} \l_{ab} |0> + {1 \over 4!} \l^a \e_{abcde} \G^{bdce} |0>$ 
where the vacuum $|0>$ contains a spinorial index.  
The condition $\l^{t} C \G^a \l = 0$ yields then that 
$\l_+ \l^a \sim \e^{abcde} \l_{bc} \l_{de}$, whereas $\l^{t} C \G_a \l = 0$ yields $\l^a \l_{ab} = 0$. 
The solution of the former equation expresses $\l^a$ in terms of $\l^+$ and $\l_{ab}$, and then 
the latter equation is also satisfied. 

According to \BerkovitsFE, physical states are defined as
cohomology classes\foot{The cohomology of the  
operator $Q$ should be computed in the infinite dimensional space of 
${x^m, \partial x^m, \partial \bar \partial x^m, \dots,  
\t^\a, \partial \t^\a, \partial \bar \partial \t^\a, \dots, \l^\a,
\partial \l^\a, \partial \bar \partial \l^\a, \dots}$, but 
one can restrict one's attention to particular 
subsets  by  using the usual filtration techniques 
(for example, the linear subspace of zero forms (functions) in (6.4)).} 
of the nilpotent BRST-like operator 
\eqn\BRST{Q_B=\oint dz \l^\a d_{z\a}} 
where $d_{z\a}$ is defined in \dd. 
Since $d_{\a}(z) d_{\b}(w)\sim - (z-w)^{-1}\g^m_{\a\b}  \Pi_m(w)$, 
\pure\ implies that $\{Q_B, Q_B\} =0$. 
The operator in \BRST\ was used in  \howe\  to show that the constraints 
of $D=(9,1),  N=1$ super-Yang-Mills theory or supergravity  can be understood as integrability 
conditions on pure spinor lines. 

Using the standard  \fms\ free OPE's 
of the worldsheet fields\foot{As usual, we combine left- and right-moving parts of the fields of the 
open string into a single field on the double interval. This
eliminates cross-terms such as $\ln(z - \bar w)$ from  
the propagators.}
\eqn\ope{p_\a(z) \t^\b(w) \sim {\d_\a^\b\over{z-w}}, 
\quad x^m(z) x^n(w)\sim -\eta^{mn} 
\log(z-w)\,,}
one obtains the following BRST transformation rules 
\eqn\BRSTtran{ [Q_B, x^m] = {1\over 2} \l^\a \g^m_{\a\b} \t^\b \,, ~~~
\{ Q_B, \t^\a \} =\l^\a\,, ~~~~~ [Q_B, \l^\a] = 0\,.} 
From these transformation rules the nilpotency of $Q_B$ when \pure\ holds is obvious. 
%on $x^m$ can also be checked directly
%$$[\{Q_B,Q_B\}, x^m] = 2 \{ Q_B, [Q_B, x^m] \} = 
%\{ Q_B, \l^\a \g^m_{\a\b} \t^\b \} = \l^\a \g^m_{\a\b} \l^\b = 0\,.$$
For later use, we also need the variation of $\Pi^m_z$
\eqn\BRSTpp{
 [Q_B, \Pi^m] = 
[Q_B, ( \partial x^m + {1 \over 2} \t^a \g^m_{\a \b} \partial \t^\b
)]= \l^\a \g^m_{\a\b} \partial \t^b\,.}  
Since the conjugate momenta $p_\a$ are independent of the variables $\t^\a$, we must also 
determine their BRST variations 
\eqn\BRSTconiu{ \{Q_B, p_\a\} = -{1 \over 2} \partial x_m \g^m_{\a \b} \l^\b - {3 \over 8} 
\g^{m}_{\a\b} \partial \t^\b  
\l^\d \g_{m \d\g} \t^\g - {1 \over 8} \g^{m}_{\a\b} \t^\b  
\partial \l^\d \g_{m \d\g} \t^\g\,, } 
Since $p_\a$ is related to $d_\a$ (see eq. \dd), it is natural to use
the latter as the fundamental variable,  
since its variation has a simpler form 
\eqn\BRSTd{ \{Q_B, d_\a\} = - \Pi_m \g^m_{\a \b} \l^\b \,.} 
It is also natural to require compatibility of the supersymmetry transformations \defsusy\ 
with the symmetry generated by $Q_B$.
Defining $[Q_\epsilon, \l^\a] = 0$, we have $\[Q_\epsilon, Q_B \] = 0$. This concludes our review 
of those results of Berkovits's program which we need. 

The BRST variation of $x_m$ given in \BRSTtran\ is not nilpotent
if $\l$ does not satisfy any constraints. We therefore introduce a new anticommuting 
worldsheet-scalar spacetime-vector $\xi^m$, whose properties will be determined as we 
proceed, and 
define
\eqn\Qa{
[Q_B', x^m] = \xi^m + {1 \over 2} \l^\a \g^m_{\a\b} \t^\b\,. 
}
We find then $Q'_B = Q_B - \oint \xi^m \Pi_m dz$. 
Similarly, since \BRSTd\ is not nilpotent, we introduce a new commuting 
worldsheet-scalar spacetime-spinor $\chi_\a$, and define
\eqn\Qb{
\{Q_B', d_\a\} = \partial \chi_\a - \Pi_m \g^m_{\a \b} \l^\b + \xi_m  \g^m_{\a \b} \partial \t^\b\,. 
}
This adds a term $-\oint \chi_\a \p_z \t^\a \, dz$ to $Q'_B$. 
The last term in \Qb\ is induced by the extra term in \Qa. Since the BRST variation of the sum of 
the last two terms is a total derivative 
we introduced $ \partial_z \chi_\a$ instead of a 1-form $  \chi_{z \a}$. 
Requiring nilpotency of $Q_B'$ on $x_m$ and $d_{z \a}$ yields then the transformation 
laws for $\xi^m$ and $\chi_\a$
\eqn\Qc{
\{Q_B', \xi^m\} = - {1 \over 2} \l^\a \g^m_{\a \b} \l^\b\,,
~~~~~[Q_B', \chi_\a] = \xi_m \g^m_{\a \b} \l^\b\,. 
}

The basic principle of our approach is to make the $\t$ variation as simple as possible, hence 
we still define $\{Q_B', \t^\a\} = \l^\a$, 
and nilpotency on $\t^\a$ leads then to $[Q_B', \l^\a]= 0$. Nilpotency on $\xi^m$ is then obvious, 
but nilpotency on $\chi_\a$ holds also as one may check by a Fierz
transformation. The ghost numbers are  
assigned such that $Q'_B$ has ghost number $+1$. Then $x_m, \t^\a$ and $d_{z \a}$ (or $p_{z\a}$) 
have ghost number zero, while $\xi^m, \l^\a$ and $\chi_\a$ have ghost number $+1$. 
Finally, the transformation rule for $\Pi^m$ becomes 
\eqn\Qd{
[Q_B', \Pi_m]=\p \xi_m + \l^\a \g^m_{\a\b} \p \t^\b\,.
}

These modified BRST 
transformation rules can be incorporated in a modified BRST current.  
Introducing the antighosts $\b^m_{z}$, %\b_{z m} = \beta_{z m} + \half (\t \g_m \k_z)$, 
$\kappa^\a_z$, and $\beta_{z \a}$ for $\xi_m, \chi_\a$, and $\l^\a$ respectively, 
the BRST current takes the following form 
\eqn\pippoold{
 j^{B \prime}_{z} \equiv
\l^\a d_{z \a} - \xi^m \Pi_{z m} - \chi_\a \p \theta^\a - \xi^m \kappa^\a_z \g_{m \a\b} \l^\b 
-  {1\over 2} \l^\a \g^m_{\a\b} \l^\b \b_{z m}\,.
}
Using the free OPE's, the OPE of two BRST currents 
can be evaluated 
\eqn\pipponil{\eqalign{ 
&j^{B \prime}_z(z) j^{B \prime}_w(w) \sim{1 \over z-w} 
\Big( \xi^m \p_w \xi_m + \, \l^\a \p_w \chi_\a  - \, \p_w \l^\a \chi_\a \Big)\,. 
}}
The double poles cancel due to the statistics of the ghost fields. 
The nonvanishing of the OPE for $j^{B\prime}_z(z) j^{B \prime}_w(w)$ implies that
 the BRST charge is not nilpotent. Therefore, it does not make sense to study the cohomology of this 
$Q$. Another related problem is that the central charge of the stress
tensor does not vanish. Both problems  
will be solved by a new ingredient. 

The nonclosure of $Q'_B$  is not very surprising. The BRST charge has the form 
\eqn\cic{ 
Q'_B = \oint dz 
\left(   C^M(z) W_{z M}(z) - \half f^{~~~~~R}_{MN} B_{z R}(z) C^N(z) C^M(z) \right)\,, 
}
where $W_{z M}(z)$ are local generators of a (super) Kac-Moody algebra, 
$ f_{MNR}$ are the structure constants of the underlying (super) Lie algebra, 
and $C^M(z)$ and $B^M_z(z)$ are the ghosts and the antighosts, respectively, 
associated to the generators $W_{z M}(z)$. However, our algebra \csm\ has 
a central charge: identifying $W_{z M} = \{ \Pi^m_z, d_{z \a}, \p_z \t^\a \}$, 
we have 
\eqn\kac{
W_{z M}(z) W_{w N}(w) \sim{ f^{~~~~~R}_{MN}  \over z-w} \,  W_{w R}(w) 
+ { h_{MN} \over (z-w)^2} \,
}
where 
\eqn\hah{
h_{MN} = \left( \eqalign{ &- \eta^{mn} ~0 ~~~~~ 0            \cr 
                                    & 0             ~~~~~~~~~ 0 ~~~~~ \d_\a^\b  \cr 
                                     &0 ~~~~~ - \d^\a_\b ~~~ 0  
}\right)
}
and the non-trivial $f^{~~~~~R}_{MN}$ are only 
$f^{~~~m}_{\a \b} = - \g^m_{\a\b}$, and $f_{\a m \b } =- f_{m\a\b} = \g_{m \a \b}$.  The 
structure constants $f^{~~~~~R}_{MN}$ and the invariant metric $h_{MN}$ 
satisfy the usual algebraic identities 
\eqn\lie{ 
f^{~~~~~R}_{MN} f^{~~~~~Q}_{RP} + {\rm cyclic~perm's} = 0\,,~~~~~ 
f_{MP}^{~~~~~R} \, h_{RN} \pm ~ f_{NP}^{~~~~~R} \, h_{MR} = 0\,.
}
The Jacobi identities follow from the Fierz identities for the Dirac matrices. 

To construct a nilpotent BRST charge, we promote the central charge to 
an operator and we introduce a new anticommuting $b-c_z$ system associated to this operator. 
The field $b$ is a 0-form with ghost number 
$-1$ and conformal spin $0$, 
and $c_z$ is a 1-form with ghost number $1$ and conformal spin $1$. 
They have the OPE
\eqn\bc{
c_z(z) b(w)  \sim{1 \over z-w} \,. 
}
Following \kawai, but with the full field $b$ instead of only the zero mode $b_0$, 
the correct BRST current $j^B_z(z)$ is then given by\foot{Another nilpotent charge depending also 
on $b$ and $c_z$ is given by  $ Q =\oint dz \[ 
\l^\a d_{z \a} - \xi^m \Pi_{z m} - \chi_\a \p \theta^\a - \xi^m \kappa^\a_z \g_{m \a\b} \l^\b 
-  {1\over 2} \l^\a \g^m_{\a\b} \l^\b \b_{z m} 
 + c_z - \half b \left( \xi^m \p_z \xi_m - \chi_\a \p_z \l^\a + \p_z
 \chi_\a \l^\a \right) + {1\over 4} b \, \partial b \,\xi^m \l^\a 
\g_{m \a\b} \l^\b \]. $}
\eqn\newpippo{\eqalign{
 j^B_{z}& = 
\l^\a d_{z \a} - \xi^m \Pi_{z m} - \chi_\a \p \theta^\a - \xi^m \kappa^\a_z \g_{m \a\b} \l^\b 
-  {1\over 2} \l^\a \g^m_{\a\b} \l^\b \b_{z m} \cr
& + c_z - \half b \left( \xi^m \p_z \xi_m - {3\over 2} \chi_\a \p_z
\l^\a + \half \p_z \chi_\a \l^\a \right) 
- {1\over 2} \p \left( b\, \chi_\a \l^\a \right) \,. 
}}
The last term is added to make this current nilpotent
\eqn\nilOPE{
j_z^B(z) j_w^B(w) \sim 0 \,.}
It is easy to check that the contributions to \nilOPE\ from the central charge in \kac\ are cancelled 
by the contributions to \nilOPE\ due to \bc.

Because the new BRST charge $ Q = \oint dz  j^B_{z}$ is nilpotent its cohomology is well-defined. 
We postpone the derivation of the transformation rules of the
antighosts until we discuss the BRST invariance  
of the quantum action. 

%%%%%%%%%%%%%%%%%%%%%%%%%%%%%%%%%%%%%%%%%%%%%%%%%%%%%%%%%%%%

\newsec{Action and Boundary Conditions}
 
The dynamics of the model is fixed by postulating free OPE's for all fields. It is natural 
to ask whether the free action is BRST invariant. This represents  
a consistency check between the structure of the BRST charge and the free OPE's used to 
construct it. For the closed string we impose as usual periodic boundary conditions. 
For the open string we shall in this way derive the boundary conditions on the fields. 
The proof that the action is invariant is equivalent to the statement
that the two-point Ward identities  
are satisfied at the tree graph level. 

Consider the tree level action 
\eqn\ACTf{\eqalign{ 
S&= \int d^2z 
\Big( {1\over 2} \p x^m \bar\p x_m + p_\a \bar\p \t^\a  + \hat{p}_\a \p \hat\t^\a \Big)\,.
}} 
We perform the BRST variation\foot{In the rest of the paper 
we often omit spinor indices; when needed we use 
parentheses as in $ (A\g^m B)$ to denote contraction of 
spinors indices.} of the left-movers
\eqn\ACTa{\eqalign{
[Q,S_L] = & \int d^2z \left[\left( \p \xi^m + {1\over 2} \p \l \g^m \t
+ {1\over 2}  \l \g^m \p \t\right) \bar\p x^m   
\right.\cr + & \left( \p \chi_\a - {1\over 2} \p x^m (\g_m \l)_\a +
{1\over 2} \p \xi^m (\g_m  \t)_\a  +  
\xi^m (\g_m \p \t)_\a \right. \cr 
- & \left. \left.  {1\over 8} (\g^m \t)_\a (\t \g_m \p \l) - {3\over 8}  ( \g^m \p\t)_\a (\t \g_m \l) 
\right) \bar\p \t^\a - p_\a \bar \p \l^\a \right] \cr
= &  \int d^2z 
\left[ \left( \p x^m + {1\over 2} \t \g^m \p \t \right) \bar\p \xi_m +
\p \t^\a \bar\p \chi_\a \right. \cr - & \left.    
\left( p_\a - {1\over 2} \p x^m (\g_m \t)_\a - {1\over 8} (\g^m \t)_\a
(\t \g_m \p \t) \right) \bar\p \lambda^\a  
\right]\,.  
}}
To obtain this result we partially integrated the BRST variation of the action 
to bring the $\bar\p$ derivatives on $\xi^m$, $\lambda^\a$ and $\chi_\a$, because then the variations 
can be canceled by  suitable BRST variations of $\beta_m$, $\beta_\a$ and $\kappa^\a$. 

The boundary terms which we produce in this way are
\eqn\buoBRST{\eqalign{
&   \half \p \Big( \xi^m \bar\p x_m \Big) 
- \half \bar\p \Big( \xi^m \p x_m \Big) 
+ {1\over 4} \p \Big( ( \l \g^m \t) \bar\p x_m \Big) \cr
&- {1\over 4} \bar\p \Big( ( \l \g^m \t) \p x_m \Big) 
+ \half\p\Big( \xi^m (\t \g_m \bar\p \t) \Big) 
- \half\bar\p\Big( \xi^m (\t \g_m \p \t) \Big) \cr
&- {1\over 8} \p \Big( (\bar\p \t \g^m \t) (\t \g_m \l) \Big) 
+ {1\over 8} \bar\p \Big( (\p \t \g^m \t) (\t \g_m \l) \Big)\,.
}}
To these terms we should add the corresponding terms from the
antiholomorphic sector. Next, we replace the total derivatives $\p$ and
$\bar\p$ by $\half \p_\sigma$ and find then the following conditions
at $\sigma = 0, \pi$ for the BRST symmetry of the open string
\eqn\bouopenBRST{\eqalign{
\xi^m \p_\tau x_m &= \hat \xi^m \p_\tau x_m \,, \cr
\chi_\a  \p_\tau \t^\a &= \hat\chi_\a  \p_\tau \hat \t^\a \,, \cr
\l \g^m \t &= \hat\l \g^m \hat \t\,, \cr
( \p_\tau \t \g^m \t ) ( \t \g_m \l) &= ( \p_\tau \hat\t \g^m \hat\t ) ( \hat\t \g_m \hat\l)\,.
 }}
A solution of all these boundary conditions, assuming that $\t^\a = \hat\t^a$, is 
\eqn\sol{
\t^\a = \hat\t^a\,, ~~~~~\xi^m = \hat\xi^m\,,~~~~ \chi_\a =
 \hat\chi_\a \,,~~~ \lambda^\a = \hat\l^\a\,, ~~~  
\e^\a = \hat\e^\a \,.
}

The BRST variation of the classical action in \ACTa\ 
can be written  in a simpler form if we use the conjugate 
momenta $\Pi^m$ and $d_\a$,  
\eqn\ACTaa{\eqalign{
[Q,S_L] =  \int d^2z \left( \Pi^m \bar\p \xi_m  + \p \t^\a \bar\p \chi_\a - d_\a  \bar\p \lambda^\a
 \right)\,.
}}
Since it is proportional 
to the equations of motions of the ghost fields, 
it can be compensated by adding a ghost-antighost action 
\eqn\ACTb{S_{L, \rm ghost} = 
\int d^2z \left( \beta_{z m} \bar \p \xi^m + \b_{z \a} \bar\p \l^\a +
 \k^\a_z \bar\p \chi_\a  + c_z \, \bar\p b \right)\,,} 
and defining the variations of the antighost fields $\beta_{z m},  \b_{z \a},  \k^\a_z$ and $b$ in a 
suitable way. These transformation rules contain various terms
bilinear in the ghosts and antighosts, needed  
to cancel the variations of the ghosts in \ACTb.  An easy way to
obtain the transformation rules of the antighosts  
is to use the BRST charge as we now discuss. 

From the BRST charge we derive the following transformation rules for the antighost fields
\eqn\ACTd{\eqalign{
&\{ Q, b\} = 1\,, \cr
&\{ Q, \b^m_z \} = - \Pi^m_z - \k_z \g^m \l + b \, \p_z \xi^m + \half
\left( \p_z b \right) \xi^m \,, \cr 
&[Q, \k^\a_z] = - \p_z \t^\a   + b\, \p_z \l^\a + {1\over 4} \left(  \p_z b\right)  \l^\a \,, \cr
&[Q, \b_{z \a}] = d_{z\a} - \b^m_{z} (\g_m \l)_\a - \xi^m (\g_m \k_z)_\a - b \, \p_z \chi_\a 
- {3\over 4} \left( \p_z b \right) \chi_\a\,.
}} 
For the new ghost $c_z$, we find
\eqn\ACTe{
\{ Q, c_z \} =  - \half \left( \xi^m \p_z \xi_m - {3\over 2} \chi_\a
\p_z \l^\a + \half \p_z \chi_\a \l^\a \right)\,. 
}
Using again integration by parts and Fierz identities, it is easy to show that 
the action is BRST invariant
$$
\Big[ Q, S_L + S_R + S_{L, \rm ghost} + S_{R, \rm ghost} \Big] = 0 \,.
$$

In the case of the open string, taking \sol\ into account, 
the last manipulations  lead to the following boundary terms 
\eqn\ghoboud{
\p_\s \left[ (b-\hat{b}) \left( \xi^m \p_\tau  \xi^m - {3\over 2} \chi_\a \p_\tau \lambda^\a + 
\half \lambda^\a \p_\tau \chi_\a \right) \right] 
}
whose solution is given by $b = \hat b$ at $\sigma = 0, \pi$. 

To derive the equations of motion of the antighost fields $\beta^\m_z, \beta_\a$ and $\kappa^\a_z$, 
one has to integrate the action by parts. This implies  new boundary
conditions on the antighost fields \lrp\  
which take the form
\eqn\bouanti{
\beta^m_\s =  \hat \beta^m_\s \,,~~~~\beta_{\s \a} =  \hat \beta_{\s \a}\,,
~~~~ \kappa^\a_\s =  \hat \kappa^\a_\s  \,,~~~~
c_\s  =  \hat c_\s\,, ~~~ {\rm at}~~~ \s = 0, \pi\,.
}

Notice that since all ghost fields are supersymmetrically invariant, the new terms in the action 
do not spoil the invariance of the theory under supersymmetry transformations.
However, for the open string we must make sure that the sum of the boundary terms, which are 
produced by susy variations, cancels \lrp. For susy we find the following boundary terms 
%($ S = \int d^2z \, {\cal L}$) 
\eqn\boususy{\eqalign{ \delta_\e {\cal L} & = 
\half \p \left[ - \half (\e \g^m \t) \bar\p x_m \right] + 
\half \bar\p \left[  \half (\e \g^m \t) \p x_m \right] \cr 
& + \half \p \left[ - \half (\hat\e \g^m \hat\t) \bar\p x_m \right] + 
\half \bar\p \left[  \half (\hat\e \g^m \hat\t) \p x_m \right] \,,
}}
and further terms with three $\t$'s. Replacing the total derivative $\p$ and $\bar\p$ by 
$\half \p_\sigma$ one finds the combination
\eqn\combsusy{
\p_\s \left[ (\e \g^m \t) \p_\tau x_m - (\hat\e \g^m \hat\t) \p_\tau x_m \right] \,.
}
Since $\p_\tau x_m$ does not satisfy any conditions, we find as condition for susy of the open string 
\eqn\bouopen{ 
(\e \g^m \t) =  (\hat\e \g^m \hat\t) ~~ {\rm at}~~~ \sigma = 0, \pi\,.
}
Similarly, the terms with three $\t$'s yield 
\eqn\consusy{ 
( \p \t \g^m \bar\p \t) (\e \g_m \t) = ( \p \hat\t \g^m \bar\p \hat\t) (\hat\e \g_m \hat\t)\,.
}
The solution of these conditions is $\e = \hat\e$. 
 
%In fact, we have \eqn\ACTc{\eqalign{[Q,S_1] = & \int d^2z \left( [Q,
%      \beta_{z m}] \bar\p \xi^m - \beta_{z m} \left( - \l \g^m \bar\p
%        \l \right) \right. \cr + & \left. [Q, \b_{z \a}] \bar\p \l^\a
%      + [Q, \kappa^\a_z] \bar\p \chi_\a + \k^\a_z \left( \bar\p \xi^m
%        \g_m \l + \xi^m \g_m \bar\p\l\right)_\a \right) }}

%%%%%%%%%%%%%%%%%%%%%%%%%%%%%%%%%%%%%%%%%%%%%%%%%%%

\newsec{The Energy-Momentum Tensor, the field $B_{zz}$  and the Ghost Current}

The antighost $b_{zz}$ for world-sheet diffeomorphisms of the bosonic
string and the RNS version of the superstring plays an important role
in the construction of higher-loop amplitudes and in the
parametrization of moduli of Riemann surfaces. We believe, therefore,
that also for the superstring a composite field $B_{zz}$, constructed
from the fundamental fields of the world-sheet theory and playing the
same role as the fundamental $b$-field, must exist. In addition, in
order to maintain an explicit covariant formulation, this composite
field $B_{zz}$ should be invariant under super-Poincar\'e
transformations of the target space.

Since the theory is a free conformal field theory, it is easy to 
write down the proper energy-momentum tensor. It follows from the action and is given by the formula
\eqn\str{\eqalign{
T_{zz} = -\half \Pi^m_z  \Pi_{m z} - d_{z \a} \p_z \t^\a - \b_{z m} \p_z \xi^m - \b_{z \a} \p_z \l^\a - 
\k^\a_z \p_z \chi_\a + \p_z b\, c_z \,.
}}
This current $T_{zz}$ is BRST invariant, as can be directly  verified
by using the BRST transformation rules  
of the fields (there are no double contractions between $Q$ and two fields in \str). 
The fact that $T_{zz}$ is BRST invariant is encouraging because the composite 
field $B_{zz}$ should satisfy the 
relation $T_{zz} = \{Q, B_{zz}\} $. 
The first two terms in $T_{zz}$ can be rewritten as 
$ -\half \p_z x^m  \p_z x^m - p_{z \a} \p_z \t^\a$,   as follows from the definition of 
$\Pi^m$ and $d_{z\a}$. In the case of open strings, $T_{zz}$ should share boundary conditions 
with the antiholomorphic partner $\bar{T}_{\bar z \bar z}$. Due to the boundary 
conditions \sol,  \ghoboud\ and \bouanti, $T$ and $\bar T$ can be
combined into a single expression which  
is defined in the whole complex $z$-plane. 

One can compute the central charge in the OPE of  $T_{zz}$ with itself, using the free OPE's 
\eqn\cencha{\eqalign{
T_{zz}(z) T_{ww}(w)  \sim
{20 \over (z-w)^4} +  {2 \over (z-w)^2} T_{ww}(w) + { 1 \over (z-w)} \p_w T_{ww}(w) + \dots }
}
The value of the central charge is obtained by summing all the 
contributions:
$(+10\times 1)_x + (-16 \times 2 )_{p,\t}$ $+ (-10 \times 2)_{\xi,\beta} + 
(+16 \times 2)_{\l,\b}$ $+ (+16 \times 2)_{\chi,\k} + (-1 \times
2)_{b,c} = 20$. Here $(n \times m)_{A,B}$  
denotes the contribution from an $A,B$ system, with $ n$  the number of components 
of the fields, and $m$ 
the central charge for a single field. Relative signs are given by the statistics of the fields. 

The resulting central charge does not vanish and, therefore, the conformal symmetry is lost. 
However,  we can compensate the non-vanishing central charge by an additional 
anticommuting spacetime-vector spin $0$ -spin $1$ system $\omega_z^m, \eta^m$. The 
final energy momentum tensor is then given by
\eqn\newT{
T_{zz} \rightarrow T_{zz} +  \p_z \eta_m \, \omega^m_z \,.
}
Due to the contribution from the $\omega_z^m, \eta^m$ system the central charge now 
vanishes. In order not to spoil the rest of the construction, we assume that these new fields 
are inert under the BRST symmetry and supersymmetry. In the following, 
we will argue that these new fields are needed for maintaining complete covariance, and 
in the next section we show how they appear when one relaxes the pure spinor constraint. 

Another important operator in the construction of the superstring 
amplitudes is the ghost current. In the present formalism, we have the following 
expression
\eqn\ghocharge{\eqalign
{
J^{gh}_z = - \left( 
\beta_{m z} \xi^m + \kappa_z^\a \chi_\a + \beta_{z \a} \l^\a + b\, c_z
+ \eta^m \omega_{z m} \right) \,,  
}}
which satisfies
\eqn\ghos{
j^B_z(z)  J^{gh}_z(w)  \sim {- j^B_z(w) \over (z-w)}  
}
Using the free OPE's one can compute the anomaly in the ghost current $J_z $ and obtains
\eqn\Jano{
J^{gh}_z(z) J^{gh}_w(w) \sim {c/3 \over (z-w)^2} = {- 11 \over (z-w)^2 }
}
The anomaly is obtained summing all the contributions: 
$(+ 10)_{\xi,\beta} + (-16)_{\l,\b}$ 
$+ (-16)_{\chi,\k} + (+1)_{b,c} + (+10)_{\eta\omega} = - 11$.

Following the idea that the theory should be described by a topological field theory 
in two dimensions, see the appendix, we only need the BRST charge, the energy-momentum 
tensor, a $U(1)$ charge and, finally, an antighost 
field $B_{zz}$ to construct amplitudes (for a pe\-da\-gogical explanation see \bvw). 
The last one is a spin-2 worldsheet-tensor (spacetime-scalar) with ghost number $-1$. 
It should satisfy the condition that it has no singular OPE with
itself \wichen\ and 
$T_{zz} = \{Q, B_{zz}\}\,$. In addition, to obtain a complete twisted $N=2$ 
superconformal algebra, the OPE with the BRST current should be given by
\eqn\Nsuper{
 j^B_{z}(z) B_{ww}(w) \sim{2\, c/ 3 \over (z-w)^3} + {2 \, J^{gh}(w) \over (z-w)^2} 
+ {2 \, T(w) \over z- w}\,}
where $T$ is given in \str\ and in \cencha, the ghost current $ J^{gh}(w) $ is given in 
\ghocharge, and $2\,c /3 = - 22$. 

Notice that in our framework $B_{zz}$ is a composite field, in contrast to the 
RNS formulation.  The requirement that $B_{zz}$ has no singular terms with itself, and 
the requirement that $T_{zz} = \{Q, B_{zz}\}$, put severe restrictions on the possible 
solutions for $B_{zz}$. 

Since the $b$ field is a Grassmann variable, we can decompose the composite operator $B_{zz}$ into 
\eqn\BBB{
B_{zz} = B^{(-1)}_{zz} + b\,  B^{(0)}_{zz} + \p_z b \, {B}^{(0)}_z +
\p^2_z b B^{(0)} + b\,  \p_z b \,  B^{(1)}_{z} \,, } 
where the upper index denotes the ghost charge of $B_{zz}$.
The condition $T_{zz} = \{Q, B_{zz}\}$ leads to a solution  
\eqn\Bparts{\eqalign{
 & B^{(-1)}_{zz}  =  \left(  \beta^m_z \Pi_{m z} - \beta_{z\a} \p_z
 \t^\a - \k^\a_z d_{\a z} \right) \,, \cr 
 & B^{(0)}_{zz}     = \half \Pi^m_z  \Pi_{m z} + d_{z\a} \p_z \t^\a -
 \p_z \eta_m \, \omega^m_z\,, \cr     
 & B^{(0)}_z      =  \half \left( - \beta^m_z \xi_m  - {1 \over 2} \beta_{z\a} \l^\a - 
                                      {3\over 2} \k^\a_z \chi_\a \right)  \,, \cr
 & B^{(0)} = 0 \,, \cr
 & B^{(1)}_{z}      =  c_z \,,
      }}
This proposal for $B_{zz}$ is, however, not nilpotent. 

The requirement that  $B_{zz}$ satisfies $T_{zz} = \{Q, B_{zz}\}$ is
not enough to fix $B_{zz}$  completely.  
In fact, one can always add a trivial part 
$B_{zz} \rightarrow  B_{zz} + \[Q, X_{zz}\]$ where $X_{zz}$ is a local polynomial with 
ghost number $-2$. Using this freedom, we construct other expressions 
that enjoy different algebraic properties. Finally, we show that one particular realization 
of $B_{zz}$ corresponds to the usual BRST charge for the bosonic string (there are no world-sheet 
fermions in our theory). 

For instance, another expression for  $B_{zz}$ is given by  
\eqn\BBBB{
B_{zz} = B^{(-1)}_{zz} + b \, {B}^{(0)}_{zz} + b\,  \p_z b  \, B^{(1)}_{z} \,, 
}
where now 
 \eqn\BBparts{\eqalign{
     &   B^{(-1)}_{zz}  = \half \left( \beta^m_z \Pi_{m z} -
     \beta_{z\a} \p_z \t^\a - \k^\a_z d_{\a z} \right) \,, \cr 
      &  B^{(0)}_{zz}     = - \half \Big( \b^m_z \p_z \xi_m + \b_\a \p_z \l^a + \k^\a \p_z \chi_a 
                                                          - 2 \, \p_z
							  \eta_m \,
							  \omega^m_z
							  \Big)\,, \cr     
 & {B}^{(1)}_z      = - {1 \over 4} \left( \Pi^m_z \xi_m  - {1 \over 2} d_{z\a} \l^\a + 
                                      {3\over 2}  \p_z\t^\a_z \chi_\a \right) + c_z \,. 
        }}
This expression again satisfies  $T_{zz} = \{Q, B_{zz}\}$. 
It is easy to check that the OPE with the BRST current $j^B_z(z)$ produces double poles 
which are field dependent. The bilinear part of these field-dependent double poles coincides 
with the ghost current \ghocharge. However, the other singularities in this OPE shows that 
the $B_{zz}$ in \BBBB\ and in \BBparts\ has neither the OPE in \Nsuper\ with $j^B_z(z)$ nor 
is nilpotent. 

A third proposal for $B_{zz}$ has much better properties than the previous two proposals. 
It differs again from the previous ones by trivial BRST terms and is given by 
\eqn\BBibbo{
B_{zz}(z) = b T_{zz}(z) + \p_z b \, \Big( J^{gh}_z(z) + b \, j^B_z(z) \Big) + \a \p^2_z b\,,
}
It satisfies the OPE in \Nsuper\ with $c=0$. Moreover it is a primary field for $\a = 35/2$
However, it is still not nilpotent since it has a singular OPE with itself. Now observe that 
the terms proportional to $\p_z b$ can be written as the BRST variation of 
$ b\, J_z(z)$ using \ghos. Furthermore, $\p_z b$ is BRST invariant. 
Therefore, we can eliminate these 
terms from $B_{zz}(z)$ to arrive to a remarkably simple expression. 
Consider the following proposal for $B_{zz}$
\eqn\Bsimple{
B_{zz}(z) = b T_{zz}(z) + \a \p^2_z b\,. 
}
This new composite operator satisfies $T_{zz}(z) = \{ Q, B_{zz}(z)\}$, since 
$\{Q, b\}=1$. Computing the OPE's 
of $B_{zz}$ 
with the ghost charge and with the energy-momentum tensor, one can check that it 
is a ghost number $-1$, anticommuting primary field with spin 2 if $\a = -\half$. By extracting the 
term proportional to $\p_z b c_z$ in $T_{zz}$, we 
end up with the expression
\eqn\Bsimple{
B_{zz}(z) = b \, \hat{T}_{zz}(z)  + b\p_z b c_z - \half \p^2_z b\,,
}
where $ \hat{T}_{zz}(z)$ is given by \str\ without the last term. 
The new field $B_{zz}(z)$ satisfies 
\eqn\Bnil{
B_{zz}(z) B_{ww}(w) \sim {2\over z-w} (\p^2_w b \, \p_w b)(w)
}
where the r.h.s. can be interpreted as an anomaly (just like the 
nonclosure term $\p c^z \p^2 c^z /(z-w)$ in the 
product of two BRST currents in RNS approach is an a\-no\-ma\-ly).  
The operator $\oint  B_{zz}(z) dz$ has the same structure (with $c \rightarrow b$)  
as the BRST charge of a bosonic 
string whose matter part is represented not only by the complete set of superstring 
matter fields $x^m, d_{z\a}$ and $\t^\a$ , but also by the ghost fields $\xi^m, \lambda^a$ and 
$\chi_\a, \omega_z^m$. To complete the correspondence with the usual bosonic BRST 
charge, one has to view the fields $b, c_z$ as the diffeomorphism ghosts $c^z, b_{zz}$ of 
the usual formulation. This follows from twisting, as we now explain. 

The $b,c_z$ system which we introduced to take care of the central charge in the Kac-Moody algebra 
is related to the diffeomorphism ghosts  $b_{zz}, c^z$ of the RNS by twisting twice: once by 
$T_{bc} = \p_z b c_z \rightarrow T' = T - \half \p_z j_z$ where $j_z = -b c_z$ which gives 
$c_z \rightarrow \psi_{3\over  2}$ and $b \rightarrow \psi_{-{1\over2}}$, and once more 
by $T'' \rightarrow T' - \half \p_z j'_z$ with $j' = - \psi_{-{1\over 2}} \psi_{3\over2}$ which 
gives $\psi_{-{1\over 2}}\rightarrow c^z$ and $ \psi_{3\over 2}\rightarrow b_{zz}$. Applying the 
twisting $b \rightarrow c^z$ and $c_z \rightarrow b_{zz}$ to the field $B_{zz}$ in \Bsimple, 
one obtains the usual BRST charge of the bosonic string, except that
the central charge of $\hat T_{zz}$ is $+2$  
instead of $+26$ and the coefficient of $\p^2 c^z$ is $-1/2$ instead of $-3/2$. 

 Conversely, our BRST charge is unitarily 
equivalent to $\oint dz \, c_z$ which is mapped, by twisting, into $\oint
dz \, b_{zz}$. Hence, our $\oint B_{zz}dz$  
and $Q$ are mapped under twisting into $Q_{RNS}$ and $\oint dz\,
b_{zz}$, respectively. On the other hand,  
in our approach, the charge $\oint B_{zz} dz$ should contribute to the
measure for higher-loop calculations  
\BerkovitsUS\ in a similar way as $\oint dz\, b_{zz}$ is used in the RNS approach. 

As is well known, in the RNS formulation (or in the bosonic string), 
the vertex operators for the open string are defined as the cohomological classes of the 
BRST operator. They can be separated into unintegrated ${\cal U}^{(1)}$ (with ghost number $+1$) 
and integrated ones $\oint {\cal V}^{(0)}_z dz$ (with ghost number
$0$) which satisfy the equations 
\eqn\desc{
\[Q, {\cal U}^{(1)}\] = 0\,,~~~~~~  \[Q, {\cal V}^{(0)}_z\] = \p_z \, {\cal U}^{(1)}\,. 
}
Defining a new operator $\delta_z$ such that $\p_z = [ Q, \delta_z]$, one has: 
$ [Q, {\cal V}^{(0)}_z] = \Big[ [ Q, \delta_z] , {\cal U}^{(1)} \Big]
= \Big[ Q, [\delta_z, {\cal U}^{(1)}] \Big]$, where  
the first equation has been used. This leads to  
\eqn\descnew{
\Big[Q, {\cal V}^{(0)}_z -  [\delta_z, {\cal U}^{(1)}] \Big] = 0  \,,
}
whose solution is ${\cal V}^{(0)}_z = [\delta_z, {\cal U}^{(1)}] + [Q,
{\cal N}^{(-1)}_z]$. Indeed, there  
is no nontrivial cohomology in this sector because $Q$ can be mapped
into $\oint dz\, c_z$ by a similarity  
transformation, see the section on comments, and the solution of $\oint dz\, c_z |\psi> = 0$ is 
$|\psi> = \oint c_z |\phi>$. This means 
that given an unintegrated vertex ${\cal U}^{(1)}$, one is able to construct the 
integrated one, namely ${\cal V}^{(0)}_z$, by acting with $\delta_z$, up to a BRST trivial part, 
denoted here by ${\cal N}^{(-1)}_z$. 
By simple manipulations, it is easy to verify that ${\delta_z = \oint dz \, B_{zz}}$. Hence
\eqn\Vertex{
{\cal V}^{(0)}_z = \oint dz B_{zz} \, {\cal U}^{(1)} + \{Q, N^{(-1)}_{z}\}\,,
 }
and we can directly check that $[Q, \oint dz {\cal V}^{(0)}_z] =
 0$~\foot{Recall that in the RNS approach  
$[Q, \left( z \, b_{-1} - b_0 \right) {\cal V}_z(z) ] = [Q, \oint dw \left( (z-w) \, b(w) \right) 
{\cal V}_z(z) ]$ and act with $Q$ on $b(w)$ and ${\cal V}_z(z)$. Double contractions from  
$Q$ to $b(w)$ and ${\cal V}_z(z)$ vanish. One obtains 
$ - {\cal V}_z(z) + b_{-1} {\cal U} + \p \left( b_0 {\cal U} - z b_{-1} {\cal U} \right)$. This yields 
${\cal V}_z(z)  = b_{-1} {\cal U}  - [ Q,  \left( z \, b_{-1} - b_0 \right) {\cal V}_z(z) ] - 
\p \left[   \left( z \, b_{-1} - b_0 \right) {\cal U}(z) \right]
$ which agrees with  our \Vertex.}.
This confirms that the field $B_{zz}$ constructed out the fundamental fields of the superstring 
plays the same role as the RNS $b$-fields, and therefore it can be used to 
fix the moduli of the Riemann surfaces. 

\newsec{Relation with the RNS superstring}

In the previous section we suggested an indentification of the ghost $b,c_z$ with the 
ghosts $c^z, b_{zz}$ of the RNS approach.  
In the present section, we will provide further relations between the RNS superstring and the present 
formulation. Some aspects of this problem have been already discussed
in  
\BerkovitsUS\ and \BerkovitsZY. In the following, we decompose $SO(9,1)$ 
spinors and vectors with respect to the $U(5)$-like subgroup as follows
\eqn\decoU{
\t^\a = \Big(\t_+, \t^a, \t_{[ab]}\Big)\,, ~~~~
p_\a =  \Big(p^+, p_a, p^{[ab]}\Big)\,, ~~~~
x^m = \Big( x^a, x_a \Big)\,, 
}
where $\t_+$ and $p^+$ are opposite-chirality $U(5)$ singlets, $ \t^a$ and $x^a$ are 
vectors in the $\underline{5}$, and $ p_a$ and $ x_a$ are vectors in the $\underline{5}^*$. The 
components $\t_{[ab]}$ and $p^{[ab]}$ are tensors in the $\underline{10}$ and 
$\underline{10}^*$, respectively. 

{\it Step 0:} We start from the RNS fields $(x^m, \psi^m, b, c, \beta, \gamma)$, where $(b,c)$ are the 
conventional diffeomorphism ghosts and $(\beta,\gamma)$ are
their superpartners. In total, we have  
12 bosonic variables and 12 fermionic ones. As shown in  \BerkovitsZY, these 
fields can be mapped into a subset of the Green-Schwarz variables:
$(x^m,  \t_+, p^+, \t^a, p_a, \b^+, \l_+)$.
%, %where $ \t_+$ and $\t^a$ are the first six component of a Green-Schwarz variable $\t^\a$. 
The opposite chirality fields $p^+, p_a$, describe the six momenta  conjugate to 
$ \t_+, \t^\a$. The two chiral bosons $\b^+, \l_+$ 
take into account the contribution of the $(\beta,\gamma)$ system, while $p^+, \t_+$ take into account 
$(b_{zz}, c^z)$. The worldsheet 
RNS fermions $\psi^m$ are decomposed into $\psi_a, \psi^a$ and 
are mapped into the spacetime fermions  $p_a, \t^a$. We should point out that this step requires 
intermediate bosonization and fermionization of $(b,c)$ and of $(\beta,\gamma)$ 
(see also \bvw\ for the D=6 case and \bv\ \fourreview\ for Calabi-Yau and $K3$ compactifications), 
with $b^+ = e^t$ and $\l_+ = e^s$.  

{\it Step 1:} In order to form a complete Green-Schwarz fermion $\t^\a$, we have to add 
the tensor components $\t_{[ab]}$ and their conjugated momenta $p^{[ab]}$. This can be done in 
a ``topological'' way, by adding a {\it BRST quartet} consisting of
the tensors $\t_{[ab]}$ and $p^{[ab]}$, and  
containing a further set of tensor fields $(v^{[ab]}, u_{[ab]})$ with opposite statistic. 
By topological we mean that the BRST charge associated to them is given by 
\eqn\BRSTquartet{ Q =  \oint dz \, u_{[ab]} p^{[ab]}\,, }
which implies that, postulating free OPE's for all the systems, $\{Q, \t _{[ab]}\} = u_{[ab]}$ and 
$[Q, v^{[ab]}] = p^{[ab]}$. Then nilpotency requires $[Q, u_{[ab]}] = 0$ and 
$\{Q,  p^{[ab]}\} = 0$. 

Together with the chiral boson $s$, the spinor $\l^\a = (e^s, u^a, u_{[ab]})$ forms a pure 
spinor (see \pure) if $ u^a = - {1\over 8} e^{-s} \e^{abcde} u_{[bc]} u_{[de]}$  \BerkovitsUS. 
The conjugate momenta $\beta_\a = (e^t, 0, v^{[ab]})$ are identified
with the eleven independent components  
of the pure spinor $\l^\a$. Counting the degrees of freedom, the number of bosons minus fermions 
cancels again as in RNS superstring. The present setting coincides with the pure spinor formulation 
\BerkovitsFE, the BRST charge is nilpotent and the Lorentz algebra can be computed in terms of 
covariant combinations of pure spinors. 

{\it Step 2:} In order to arrive at a completely covariant formulation, we let
all the components of the spinor $\l^\a$ become independent. Namely we want to remove the pure spinor 
constraint by introducing  a new independent field $u^a$ and its conjugate momentum $\beta_a$ 
in order to reconstruct a complete spinor. Again, we introduce a BRST trivial quartet consisting of 
the spinor parts $ u^a$ and $ v_a$, and half of a spacetime vector $\xi^a$ and 
its conjugate momentum $\beta_a$. Notice that the statistics of the new fields should be opposite to 
those of $\beta_\a$ and $\lambda^\a$, and, therefore, $\xi^a$ and $\beta_a$ are anticommuting 
vectors. 

We can also understand the necessity of the new degrees of freedom by observing that, relaxing 
the pure spinor constraint $\l \g^a \l = 0 $  is equivalent to set 
\eqn\Vexi{
\{ Q, \xi^m \} = - \half \l \g^m  \l \,.
}
(Recall that $\l \g_a \l = 0$ is automatically satisfied if 
$\l \g^a \l =0$ is satisfied). The $\underline{5}$ part of this vectorial equation 
can be solved in terms of $u^a$:
\eqn\Vexo{
u^a = e^{-s} \left(  \{ Q, \xi^a \}  -  
{1\over 8} \e^{abcde} u_{[bc]} u_{[de]}\right)\,.
}
Then, finally substituting $u^a$ into the $\underline{5}^*$ part of
equation \Vexi, one gets $\{ Q, \xi_a \} =  
  e^{-s} \{ Q, \xi^b \}  u_{[ba]}$ which implies that the vector $\xi^m$ is a constrained 
vector. The new fields $ u^a, v_a$ and $\xi^a, \beta_a$  form again a BRST quartet. 

{\it Step 3:} To remove the vectorial constraint, we can further enlarge the 
field space by introducing another quartet formed by the missing components of the 
vector $(\xi_a, \beta^a)$ (removing the constraint) and introducing a new vector 
$\chi_a$ with its conjugate momentum $\kappa^a$. The latter can be viewed as the $\underline{5}$-part 
of $\chi_\a$  and of its conjugate momentum $\kappa^\a$. As above, the introduction of these 
new fields does not affect the physics of the theory. 

{\it Step 4:} All the constrains are now removed, however the spinor $\chi_\a$ and its conjugate 
momentum are treated non-covariantly. A completely covariant approach requires 11 more 
anticommuting fields. One of these new fields, namely the scalar $b-c_z$ system, 
already entered the formalism by requiring the 
nilpotency of the BRST charge following \kawai. The further 10 fields $\omega^m_z$ and $\eta_m$ 
are necessary for the vanishing of the central charge. Together with the $\underline{1}$- and 
$\underline{10}$-part of $\chi_\a$, they saturate the number of degrees of freedom to 
make the formalism covariant, preserving unitarity, the conformal invariance and 
the cohomology of the BRST operator. 

\newsec{Massless states for open strings}

The cohomology of the operator $Q_B$ in \BerkovitsNN\ -- denoted by $H^n(Q_B)$, where $n$ 
is the ghost number --  
is a constrained cohomology with the condition \pure\ for pure spinors $\l^\a$. 
Berkovits argued that $H^1(Q_B)$ must be linear in the 
field $\l^\a$, because of the nonlinearity of the action of the
Lorentz generators in the fields $\l_{ab}$. 
It yields the spectrum of the superstring
%.  For massless modes, the 
%requirement of linearity of $H^1(Q_B)$ implies the super-Poincar\'e covariant 
%equations of motion 
for SYM \SYM\ (or SUGRA for closed strings), but for massive modes, 
an explicit parametrization of $\l^\a$ seems to be needed. 

In our approach based on the Kac-Moody algebra \kac, we must find the
appropriate constraints on the kinematics and the dynamics of physical
states.  We need therefore both a kinematical constraint which relates the
curvatures to the connections (supermultiplets of $N=1$ superspace
$D=(9,1)$ in the case of open strings \SYM) and a dynamical constraint
which imposes the correct equations of motion on the connections (for
example the super Yang-Mills equations of motion in $D=(9,1)$ \SYM).  

In the conventional RNS formalism \GS\ \polc, the dynamical properties of the
physical states are encoded in the BRST symmetry which implements, at
the quantum level, the super-Virasoro algebra of the spinning string,
while the supersymmetry multiplets are selected by means of the GSO
projection on the BRST cohomology. The BRST cohomology describes the
possible deformations of the super-Virasoro algebra in terms of
unintegrated vertex operators ${\cal U}^{(1)}$ with ghost number $+1$
or, equivalently, in terms of integrated vertex operators 
$\oint dz {\cal V}_z^{(0)}$.  They are related by the equation 
\eqn\cohoA{ \left[Q,
{\cal V}^{(0)}_z \right] = \p_z \, {\cal U}^{(1)} \,, 
} 
where ${\cal V}^{(0)}_z$ has zero ghost number. This equation implies that 
$\oint dz {\cal V}_z^{(0)}$ is BRST invariant. 

In our case the BRST algebra encodes the Kac-Moody algebra and the corresponding 
BRST cohomology should contain all the possible deformations of the underlying algebra. In the 
following, we will show that indeed the ghost number one BRST cohomology of $Q$, represented 
here by the unintegrated vertex ${\cal U}^{(1)}$ with ghost number $+1$, describes all 
the admissible deformations. The constraints on the deformations are obtained as Bianchi identities 
on the curvatures of the gauge superpotentials $A_\a(x,\t), A_m(x,\t)$
and of the field strenght $A^\a$.  
This analysis does not lead to the field equations which are implemented by requiring that the possible 
deformations preserve the energy-momentum tensor $T$ in equation \str.  

First, we discuss the BRST cohomology and then we discuss the deformations which 
preserve the energy-momentum tensor. 

Consider the unintegrated vertex ${\cal U}^{(1)}$ and the computation of the cohomology 
for the operator $Q$. The physical space of vertex operators is contained in the cohomology of $Q$ 
at ghost number $1$, which means that  ${\cal U}^{(1)}$ satisfies
\eqn\cohoB{
\left[Q, {\cal U}^{(1)} \right] = 0\,, ~~~~~~     {\cal U}^{(1)} \neq
[ Q, \Omega ] \,,
} 
where $\Omega$ is a ghost number zero superfield. Notice that ${\cal U}^{(1)}$ could depend 
on the auxiliary fields $\eta^m$ and $\omega_{zm}$. Nevertheless,
since  $\eta^m$ and $\omega_{zm}$, as well  
as the current $Q_{\eta} \equiv \oint dz \, \eta^m \omega_{zm}$,  are
trivial under the BRST transformations,  
we expect that the physical observables are independent of them. The 
requirement that the physical vertex  ${\cal U}^{(1)}$ does not contain $\eta^m$ can be written as 
\eqn\neweq{
\oint \eta^m \omega_{z m} \, {\cal U}^{(1)} = 0\,. }
Since ${\cal U}^{(1)}$ is a scalar, it can not depend on $\omega_z^m$. 

To solve \cohoB, we decompose  ${\cal U}^{(1)}$ as 
\eqn\cohoC{\eqalign{
{\cal U}^{(1)} &= \l^\a A_\a + \xi^m A_m + \chi_\a A^\a \cr
& + b\, \Big( \l^\a \l^\b F_{\a\b} + \l^\a \xi^m F_{\a m} + \xi^m \xi^n F_{m n}   
+ \l^\a \chi_\b F^{~~\b}_\a + \chi_\a \, \xi^m F^{\a}_{~~m} + \chi_\a \chi_\b F^{\a \b} \Big)\,, 
}}
where $ A_\a, \dots, F^{\a \b}$ are arbitrary superfields of $x_m, \t^\a$. 
Terms with derivative $\p_z x^m$, $\p_z \t^\a$ and $d_{z\a}$ cannot be present since 
${\cal U}^{(1)}$ is a worldsheet scalar. Perhaps one may construct a
composite field $C^z$ (after suitable  
bosonization and fermionization to allow contravariant vectors) but the cohomology derived from \cohoC\ 
should not change. 
Note that, because $b$ has ghost number $-1$ and $b(z) b(w) \sim0$,  we can only include terms 
which are at most bilinear in $\xi^m, \chi_\a$ and $\l^\a$.
Acting on ${\cal U}^{(1)}$ with the BRST charge $Q$ (given in \newpippo), and collecting 
the independent contributions, we find the following equations 
\eqn\cohoD{\eqalign{ 
\l\l:~~& D_{(\a} A_{\b)} - \half \g^m_{\a\b} A_m + F_{\a\b} = 0 \,, \cr
\l\xi:~~& \p_m A_\a - D_\a A_m + \g_{m\a\b} A^\b + F_{\a m} = 0 \,, \cr
\xi\xi:~~& \p_{[m} A_{n]} + F_{mn} = 0 \,, \cr
\l\chi:~~& D_\b A^\a + F_\b^{~~\a} = 0 \,, \cr
\xi\chi:~~& \p_m A^\a + F_{~~m}^{\a} = 0 \,, \cr
\chi\chi:~~& F^{\a\b} = 0 \,, 
}}
where $D_\a \equiv \p / \p \t^\a + \half \t^\b \g^m_{\a\b} \p / \p x_m$. The normalization is chosen 
such that $D_\a D_\b + D_\b D_\a = \g^m_{\a\b} \p_m$. We define $D_{(\a} A_{\b)} = 
\half \left( D_\a A_\b + D_\b A_\a \right)$ and $\p_{[m} A_{n]} = 
\half \left( \p_m A_n - \p_n A_m \right)$. 
From the terms containing $b$, we get 
\eqn\cohoE{\eqalign{ 
\l\l\l:~~& D_{(\a } F_{\b \g) } -  \half \g^m_{(\a \b} F_{\g) m} = 0 \,, \cr
\l\l\xi:~~& \p_m F_{\a\b} - D_{(\a} F_{\b) m} - \g^m_{\a\b} F_{mn} +
\g_{m \g (\a} F_{\b)}^{~~\g} = 0 \,, \cr 
\l\xi\xi:~~& \p_{[m} F_{\a| n]} + D_\a F_{mn} - \g_{_{[m | \a \b }} F_{~~n]}^{\b} = 0 \,, \cr 
\l\l\chi:~~& D_{(\a} F_{\b)}^{~\g} - \half \g^m_{\a\b} F_{~~m}^{\g} = 0 \,, \cr 
\l\xi\chi:~~& \p_m F_\a^{~~\b} - D_\a F_{~~m}^{\b} + 2 \g_{m \a\g} F^{\b\g} = 0 \,, \cr
\xi\xi\chi:~~& - \p_{[m} F_{~~n]}^{\a} = 0 \,, \cr
\l\chi\chi:~~& D_\a F^{\b\d} = 0 \,, \cr
\xi\chi\chi:~~& \p_m F^{\a \b} = 0 \,, 
}}
These equations are written in terms of superfields, and therefore,
 supersymmetry is manifest. Equations \cohoE\ correspond to Bianchi identities for the 
curvature defined in \cohoD, so they are automatically satisfied when the curvatures are expressed 
in terms of the potentials as in \cohoD. Notice that in $D=(9,1)$-superspace, imposing 
$F_{\a\b} = F_{m\a} =0$ one gets the equations \wittwi\ 
 \eqn\neh{\eqalign{
& \g_{[mnpqr]}^{\a\b} D_\a A_\b = 0 \,, \cr
& A_m = {1\over 8} \g_m^{\a\b} D_\a A_\b \,, \cr
&A^\a = {1\over 10} \g^{m\a\b} \left( D_\b A_m - \p_m A_\b \right)\,. \cr}   
}
These equations imply this other system of equations 
\eqn\nehh{\eqalign{
& D_\a A_m - \p_m A_\a + \g_{m \a\b} A^\b = 0 \,, \cr
& D_\a A^\b = {1 \over 4} \g^{mn~\b}_{~\a} F_{mn}\,, \cr
& D_\a F_{mn} = ( \g_{m \a\b} \partial_n -  \g_{n \a\b} \partial_m ) A^\b \,.
}}
The latter, together with the Bianchi identities \cohoE, imply the
equations of motion for linearized super Yang-Mills  
\wittwi\ 
\eqn\nehhh{\eqalign{
\g^m_{\a\b} \partial_m A^\b = 0 \,, ~~~~~~~ \partial^m F_{mn} = 0 \,.  
}}
In ref \har\ it is shown (in the Wess-Zumino gauge) that \nehhh\ implies \nehh\ and 
\nehh\ implies \neh. Hence, all three set of equations of motion are equivalent. 

The potentials $A_\a, A_m$ of the 
gauge connection and their corresponding curvatures $F_{\a\b}, \dots, F^\a_{~~m}$ parametrize 
all the possible deformation of the Kac-Moody algebra. Indeed, we can observe that, if we 
define the new BRST  $Q_{\cal U} = Q +  {\cal U}^{(1)}$, the nilpotency of the new BRST charge 
implies (up to terms quadratic in the vertex $ {\cal U}^{(1)}$) equation \cohoB, 
namely $\left[Q, {\cal U}^{(1)} \right] = 0$. The new BRST charge is given by 
\eqn\newp{\eqalign{
Q_{\cal U} & = \oint dz \Big[  
\l^\a \left( d_{z \a} + A_\a \right) 
- \xi^m \left(\Pi_{z m} - A_m\right)  - \chi_\a \left(\p \theta^\a - A_\a\right) \Big. \cr
& \left. - \xi^m \kappa^\a_z \g_{m \a\b} \l^\b 
-  {1\over 2} \l^\a \g^m_{\a\b} \l^\b \b_{z m} \right.  \cr 
&   + c_z - \half b \left( \xi^m \p_z \xi_m - {3\over 2} \chi_\a \p_z
\l^\a + \half \p_z \chi_\a \l^\a \right) 
- {1\over 2} \p \left( b\, \chi_\a \l^\a \right) \cr
& \left.  + b\, \Big( \l^\a \l^\b F_{\a\b} + \l^\a \xi^m F_{\a m} + \xi^m \xi^n F_{m n}   
+ \l^\a \chi_\b F^{~~\b}_\a + \chi_\a \, \xi^m F^{\a}_{~~m} + \chi_\a \chi_\b F^{\a \b} \Big) 
\right]\,. 
}}
where the Kac-Moody generators $\Pi^m_z, \p_z \t^\a, d_{z\a}$ are shifted by the gauge potentials 
$A_m, A_\a$ and $A^\a$.  In the same way, the energy momentum tensor $T_{zz}(z)$ in equations \str\ 
is modified into a new tensor $T^A_{zz}(z)$
\eqn\strA{\eqalign{
T^A_{zz}(z) &= -\half \left( \Pi^m_z  - A_m \right) \left(\Pi_{m z} - A_m \right)  
- \left( d_{z \a} + A_\a \right) \left( \p_z \t^\a - A^\a \right) \cr
& - \b_{z m} \p_z \xi^m - \b_{z \a} \p_z \l^\a - 
\k^\a_z \p_z \chi_\a + \p_z b\, c_z + \p_z \eta^m \omega_{zm} \,.  
}}
Finally, by requiring that this $T^A_{zz}(z)$ satisfies the usual OPE, namely 
\eqn\OPEstr{\eqalign{
T^A_{zz}(z) T^A_{ww}(w) \sim {2 T^A_{ww}(w) \over (z-w)^2} + {\p_w
T^A_{ww}(w) \over  (z-w)} + {\cal O}(A^2)\,,  
}}
we find the constraints on the gauge potentials and on the field strenghts. The 
double poles yield
\eqn\fe{\eqalign{
& \p^2 A_\a + 2 \g^m_{\a\b} \p_m A^\b + D_\a D_\b A^\b - D_\a \p_m A^m = 0\,, \cr 
& \p^2 A^\a = 0\,,  \cr
& \p^2 A_m - \p_m \p_n A^n + \p_m D_\a A^\a = 0\,,
}}
and the simple poles give further equations
\eqn\fe{\eqalign{
& D_\a\left( \p^2 A_m - \p_m \p_n A^n \right) + \g^m_{\a\b} \p_m A^b + 
    \half \p_m \p^2 A_\a + \p_m D_\a D_\b A^\b = 0 \,,\cr
& \p^2 A_\a + D_\a D_\b A^\b - D_\a \p_n A^n = 0\,. 
}}
From these equations, we immediately get the equations of motion and the gauge-fixing 
\eqn\motion{\eqalign{
& \p^2 A_\a = 0\,, ~~~~~~ \g^m _{\a\b} \p_m A^\b = 0\,, ~~~~~~ \p^2 A^\a = 0 \,, \cr
& \p^2 A_m = 0 \,, ~~~~~~  \p_n A^n =  D_\a A^\a  + f\,.
}}
Here $f$ is a constant. These correspond to SYM equations of motion. 
To see this, it is convenient to expand the superfields $A_m$ and $A^\a$ to first power of 
$\t^a$ in terms of gauge field $a_m(x)$ and of gaugino $u^\a(x)$ (in
the Wess-Zumino gauge  $\t^\a A_\a =0$  \OoguriPS) 
\eqn\decAA{\eqalign{
A_m(x, \t) &= a_m(x) +  \t^a \g_{m \a\b} u^\b(x) + \dots \,, \cr
A^\a(x, \t) &= u^\a(x) +  \g^{mn,\a}_{\b} \t^\b \p_m a_n(x) + \dots \,,  
}}
where the coefficients  in front of the $\t$-terms are fixed by
supersymmetry. The equation $\p^2 A_m =0$   
implies that $\p^2 a_m(x) = 0$, while $ \g^m _{\a\b} \p_m A^\b = 0$ implies that 
$\g^m _{\a\b} \p_m u^\b(x) = 0$ at zero order in $\t$. The latter
coincides with the gaugino equation of motion.   
However, we still have a condition at the same order by means of $ \p_n A^n =  D_\a A^\a$ 
(where the constant $f$ is set to zero for convenience). By 
inserting the expansions \decAA, we deduce $\p_m a^m = 0$, hence the gauge field $a_m$ satisfies 
the usual Maxwell equation of motion in the Landau gauge. 
As shown in \har\ these equations of motion, with the Bianchi identites \cohoE\  
imply the constraints $F_{\a\b} = F_{\a m} = 0$. 

Notice that $F_{\a\b} = F_{\a m} = 0$ parametrize a particular set of all the possible deformations 
of the Kac-Moody algebra. Namely, those deformations which do not modify 
the non-abelian Lie algebra of zero modes $\oint dz \, \Pi^m_z$ and
$\oint dz \, d_{z \a}$.   Essentially, 
this means that among all possible deformations of the Kac-Moody algebra, the physical observables 
are those which do not modify the non-abelian part of the algebra. Clearly, the abelian part of the 
algebra due to $\Pi \, \Pi \sim z^{-2}\,, \p\t \, \p\t \sim 0\,, \Pi
\, \p\t \sim 0$ and $d \, \p\t \sim z^{-2}$ 
is deformed by the curvatures $F_{mn}, F_m^{~~\a}$ and $F_\a^{~~\b}$. Finally, we 
have to point out that the introduction of the $b-c_z$ system to promote the central charge of the 
Kac-Moody algebra \kac\ to a generator renders the BRST cohomology trivial (as pointed out in  
\kawai), however this parametrizes correctly all the possible
deformations of the algebra. It is only the  
requirement on the deformed energy-momentum tensor  which implies the correct equations of motion. 

%The first two equations of  \cohoD\ can be solved and yield \har\
% \eqn\neh{\eqalign{
%%& F_{\a\b} = F_{\a m} = F^{\a\b} = 0 \,, \cr 
%& \g_{[mnpqr]}^{\a\b} D_\a A_\b = 0 \,, \cr
%& A_m = {1\over 8} \g_m^{\a\b} D_\a A_\b \,, \cr
%& A^\a = {1\over 10} \g^{m\a\b} \left( D_\b A_m - \p_m A_\b \right)\,, \cr}   
%}
%Substituing the constraints on the supercurvature into the Bianchi identities in \cohoE\
%the complete solution reads
%\eqn\nehh{\eqalign{
%& D_\a A^\b = {1 \over 4} \g^{mn~\b}_{~\a} F_{mn}\,, \cr
%& D_\a F_{mn} = ( \g_{m \a\b} \partial_n -  \g_{n \a\b} \partial_m ) A^\b \,, \cr
%& D_\a D_\b A^\a = 0 \,, \cr
%& \g^m_{\a\b} \partial_m A^\b = 0 \,, \cr
%& \partial^m F_{mn} = 0 \,, \cr
%& \partial_{r} F_{mn} +   \partial_{m} F_{nr} + \partial_{n} F_{rm} = 0\,.
%}}
%It describes a SYM gauge field $A_\a$ on-shell. 
%Choosing the gauge $\t^\a A_\a = 0$, we parametrize 
%\eqn\parA{
%A_\a = a_m (\g^m \t)_\a - {2\over 3} (\g^m \t)_\a (\t \g_m \psi) + \dots
%}
%where $a_m(x)$ satisfies the linearized Maxwell equations and $\psi^\a(x)$ the 
%linearized Dirac equation. 

Using the method discussed in \har\ to eliminate the auxiliary fields from the superfield
$A_\a$, one arrives at the expressions \OoguriPS
\eqn\oo{\eqalign{
 A_\a &= (\gamma^m \t)_\a
\left[ a_m
-{2 \over 3} (\t \g_m u) 
  - {1\over 8}(\t \gamma_m \gamma^{n r}
  \t) f_{n r} + \cdots\right]\,, \cr
 A_m & =  a_m - (\t \gamma_m u) 
   - {1\over 4} (\t \gamma_m \gamma^{n r}\t)
      (f_{n r} + {2 \over 3} \gamma_{[n} \p_{r]} u)
+ \cdots\,, \cr
A^\a & =
 u^\a + {1 \over 2} (\gamma^{m n} \t)^\a
 ( f_{m n} - \t \gamma_{[ m}
  \p_{n]} u ) +\cdots \,.
}}
The superfields in \oo\ are written exclusively in terms
of the physical gauge field $a_{m}$, its field strength $f_{m n}$,
and the gaugino $u^\a$; all auxiliary fields have been eliminated.
Moreover, the gauge-fixing condition  $\t^\a A_\a = 0$ is automatically
satisfied.  

As indicated in \cohoA\  one can determine the integrated vertex ${\cal V}^{(0)}_z$ from 
the unintegrated vertex ${\cal U}^{(1)}$ by solving the equation $[ Q,
{\cal V}^{(0)}_z] = \p_z  {\cal U}^{(1)}$. We have found the complete
$ {\cal V}^{(0)}_z$ . The first few terms are the following:
\eqn\inteVERT{\eqalign{ 
{\cal V}^{(0)}_z &= \Pi^m_z A_m + \p_z \t^\a A_\a + d_{z \a} A^\a  \cr
&+ 2 \b^m_z \xi^n F_{mn} + \b_{z \a} \l^\b D_\b A^\a + \k^\a_z \chi_\b D_\a A^\b \cr 
&+ \left( \b^m_z \chi_\a - \b_{z \a} \xi^m \right) \p_m A^\a + \dots \,.
}
}
The first three terms were first proposed by Siegel \csm, while the
next three terms are the covariantization  
of the vertex obtain by Berkovits \BerkovitsFE. The last  term in \inteVERT\  
is new, and there are also further terms proportional to $b$ and $b \, \p b$
which will be published elsewhere.

%%%%%%%%%%%%%%%%%%%%%%%%%%%%%%%%%%%%%%%%%%%%%%%%%%%%%%%%%%

%%%%%%%%%%%%%%%%%%%%%%%%%%%%%%%%%%%%%%%%%%%%%%%%%%%%%%%%%%%

\newsec{Comments}
We end with some comments. 

{\it 1.} Our approach is based on a BRST charge for a Kac-Moody algebra, and it is 
not, in first instance, of the usual form $\oint dz  c^z T_{zz} + \dots$. The generators of the Kac-Moody algebra yield 
spacetime symmetries instead of worldsheet symmetries. However, by introducing a ghost field $c^z$ and 
replacing the ghost fields $\xi^m,\l^\a$ and $\chi_\a$ 
by $c^z \Pi^m_z, c^z \p_z \t^\a$ and $c^z d_{z\a}$ one recovers the
familiar form $Q = \oint c^z T_{zz} + \dots$.  

{\it 2.} The field $c^z$ plays the role of the diffeomorphism ghost, but in our 
formalism it should be a composite field. It is under construction.
With this $c^z$ one should be able to   
construct unintegrated vertex operators and prove the equivalence with the RNS formulation. 

{\it 3.} In \wie\ an action was constructed for the superstring using Maurer-Cartan equations for the 
Lie algebra containing $\Pi, \p \t, d_\a$ and a central charge. The complete  action includes a 
WZNW term. By replacing the exterior derivative $d$ by $Q + d$, one can construct 
simultaneously the action of \wie\  and 
our BRST charge and we can prove the equivalence with the conventional Green-Schwarz formalism. 
It turns out that the anomaly term in the BRST charge (the term
proportional to $b$) corresponds to the WZNW  
term in the action. Our action, which did not (yet) include a WZNW term,
is separately invariant under our  
$Q$ as we have shown. This is possible when the WZNW term is also separately invariant under our 
BRST charge. 

{\it 4.} We have found a field $B_{zz}(z)$ which yields the integrated vertex with 
${\cal V}^{(0)}_z$ from the unintegrated vertex ${\cal U}^{(1)}$. In this construction 
we found other candidates for $B_{zz}(z)$ which are related by $Q$-exact terms. Specifically, 
$B^I$ in \BBB, $B^{II}$ in \BBBB, $B^{III}$ in \BBibbo, and $B^{IV}$ in \Bsimple\ are related 
as follows 
\eqn\BBs{ B^I - B^J = Q \left[ b \left( B^I - B^J \right) \right]\,.} 

{\it 5.} It might seem that our BRST current is trivial because it can be produced 
by a similarity transformation of $c_z$: 
\eqn\sims{
e^{- \oint \half b \tilde{j}^B_z } c_z e^{\oint \half b \tilde{j}^B_z } 
= c_z +  \tilde{j}^B_z - \oint b \tilde{j}^B_z \tilde{j}^B_z \,. 
} 
The last term is the anomaly in the BRST charge and the right-hand side is indeed our 
BRST current $j^B_z$. However, we also must make this similarity transformation on $\oint B_{zz} dz$ 
and this does not yield a trivial result. These issues are intimately related to the notion of {\it 
big picture} (that is the Hilbert space of the RNS string which contains the zero mode of $\xi$, 
where $\xi$ is defined by the bosonization of RNS superghosts $\beta,\gamma$ in the 
usual way). In the RNS case, we can split the BRST charge $Q$ into 
$Q_0 + Q_1$ where $Q_1 = \oint \half b \, \gamma^2 = \oint b \,\eta \,
\p \, \eta e^{-2\, \phi}$. A similarity  
transformation of $Q_1$ with $ e^{- x Q_0}$, where $ x = c \, \xi \, \p\, \xi e^{-2 \, \phi}$, produces 
$Q$. Note that since $\xi_0$ is  present, the charge $Q$ is equivalent to the trivial charge $Q_1$ 
in the big picture. Also here, one should transform $\oint b_{zz}
dz$. Of course,  the BRST charge in the  
RNS formalism is not trivial.

%%%%%%%%%%%%%%%%%%%%%%%%%%%%%%%%%%%%%%%%%%%%%%%

\newsec{Acknowledgements}
We thank R. Stora and W. Siegel for discussions. P.A.G. thanks CERN for 
the hospitality in July 2001 where this work was started.  G.P. thanks New York University 
and Stony Brook for visits. This work was partly funded by NSF Grants PHY-0098527 and PHY-0070787. 

\newsec{Appendix: Antifields}

We will elsewhere present the BV formulation of our results. We mention here the first 
step. 

We add all the possible antifields $x^*_m, \t^{*,\a}, \l^{*,\a},$  
$d^*_\a, \xi^*_m$ and $\chi^*_\a$ (with ghost numbers $-1,-1,-2,-1,-2,-2$; notice that 
$d^*_\a$ is an antiholomorphic vector $d^*_{\bar z, \a}$ ) and 
we can construct the antifield-dependent part of the action 
from the variation of   $x^m, \t^\a, \l^\a, d_\a, \xi^m$ and $\chi_\a$
\eqn\ACT{S_{s} = \int d^2z \Big[ x^{*}_m ( \xi^m + {1\over 2} \l^\a \g^m_{\a \b} \t^\b ) + 
( -  \Pi_m \l^\a \g^m_{\a \b} - \partial \t^\a \g^m_{\a \b} \xi_m + \partial \chi_\b ) \, d^{* \b} } 
$$+ \, \l^\a \t^*_\a  - {1 \over 2} \l^\a \g^m_{\a \b} \l^\b \xi^*_m +
\xi_m \l^\a \g^m_{\a \b} \chi^{* \b} \Big]\,. $$  

The BRST symmetry is compatible with the supersymmetry
if all the fields $\xi^m, \lambda^\a$ and $\chi_\a$ are susy invariant. Furthermore, 
the action $S_s$ is supersymmetric if we define 
\eqn\susyanti{\eqalign{[Q_\epsilon, \t^*_\a] = + { 1\over 2} x^*_m \g^m_{\a\b} \epsilon^\b\,, ~~~~~~ 
[Q_\epsilon, x^*_m] = 0\,,}} 
all the other susy transformation rules for the antifields being
zero. Moreover, the combination  of antifields  
$\theta^*_\a + {1\over 2} x^*_m \g^m_{\a\b} \t^\b$ is supersymmetric. 

On the other hand, by differentiating the action $S_s$ in\ACT\ with
respect to fields, we derive the BRST variations of the antifields
\eqn\BRSTanti{\eqalign{ &\{Q'_B, x^*_m\} = \partial \Big( \l^\a
\g_{m\a\b} d^{* \b} \Big)\,, \cr &[Q'_B, \xi^*_m] = - x^*_m + \partial
\t^\a \g_{m \a\b} d^{*\b} + \l^\a \g_{m \a \b} \chi^{*\b}\,, \cr &[Q'_B,
\t^*_\a] = - {1 \over 2} x^*_m \g^m_{\a \b} \l^\b - \partial \Big(
\xi_m \g^m_{\a \b} d^{* \b} \Big) - \g^m_{\a \b} \partial \t^\b \l^\d
\g_{m \d\g} d^{* \g} - {1\over 2} \t^\b \p \Big(\l^\d \g_{m \d\g} d^{*
\g}\Big)\,, \cr &\{Q'_B, \l^*_\a\} = \t^*_\a + {1 \over 2} x^*_m \g^m_{\a
\b} \t^\b - \Pi_m \g^m_{\a \b} d^{* \b} - \xi^*_m \g^m_{\a \b} \l^\b +
\xi_m \g^m_{\a \b} \chi^{*\b}\,, \cr &\{Q'_B, \chi^{* \a}\} = - \partial
d^{* \a}\,, \cr &[Q'_B, d^{* \a}] = 0\,. }}  

There are duality relations among fields and antifields, for example the linear shift 
of $x^m$ by means of the vector $\xi^m$ corresponds to the linear shift of $\xi^*_m$
by means of $x^*_m$. In the same way, the variation of $d_\a$ contains
the holomorphic derivative of $\chi_\a$, which is dual to the
variation of $\chi^{*\a}$ into $ - \p_z d^{*,\a}$.  This duality between fields and antifields
is typical for topological field theories quantized with
the BV formalism. To reveal such topological aspects, it is
convenient to introduce new variables 
\eqn\newfields{\eqalign{ 
&\widetilde\xi^m = \xi^m + {1\over 2} \t^\a \g^m_{\a\b}\l^\b \,,\cr 
&\widetilde{d}_{z \a} = d_{z \a} + \p_z x^m \g_{m \a\b}\t^\b + {1\over
6} \g^{m}_{\a\b} \t^\b \t^\g \g_{m \g\d} \p_z 
\t^\d \,,\cr 
&\widetilde{\chi}_\a = \chi_\a + \widetilde\xi^m \g_{m\a\b}\theta^\b + 
{1\over 6}   \g^{m}_{\a\b} \t^\b \t^\g \g_{m \g\d} \l^\b \,. 
}} 
Inserting the expression for $d_{z \a}$ into \newfields, we find that
$\widetilde{d}_{z \a} = j^\e_{z \a}$  
where $ j^\e_{z \a}$ is the current for the supersymmetric charge given in \defsusy. 
In terms of the new variables the BRST transformations simplify to 
\eqn\newBRST{\eqalign{ 
&[Q, x^m] =\widetilde\xi^m\,,~~~~~~ \{Q,\widetilde\xi^m \} =0\,,\cr 
&\{Q, \t^\a\} = \l^\a\,, ~~~~~~ [Q, \l^\a] = 0\,,~~~~ \cr
&\{Q, j^\e_{z \a}\} = \partial_z \widetilde{\chi}_\a\,,~~~~ [Q,\widetilde{\chi}_\a]= 0 \,,
 }}  
which clearly show the topological nature of the BRST
symmetry generated by $Q$. Notice that although 
the BRST symmetry is very simple in terms of the new variables, 
the variables $\widetilde\xi_m, \widetilde{d}_{z\a}$ and
$\widetilde{\chi}_\a$ are no longer susy invariant.  
In terms of the old variables the BRST symmetry is rather complicated, but $\Pi_m, d_{\a}$ and all the
ghost fields are supersymmetric (clearly $x^m$ and $\t^\a$ are not
supersymmetric being the coordinates of the superspace). 

Due to the simple structure of \newBRST, we can write the associated BRST charge as
\eqn\pippo{\eqalign{ 
Q &= \oint dz j^B_{z}\,, \cr
j^B_{z} &\equiv \l^\a  \left( j^\e_{z \a}  + {\cal J}^{\e}_{z \a}
\right) - \widetilde\xi^m \, \p_z x_m +  
\p_z \hat\chi_\a \, \t^\a \,, \cr   
{\cal J}^{\e}_{z \a} &\equiv \half \widetilde\b_{z m} \g^m_{\a \b} \l^\b - 
\widetilde\xi_m \g^m_{\a\b} \widetilde\k^\b_z \,, 
 }}  
where ${\cal J}^{\e}_{z \a}$ generates the susy transformations on the 
ghost fields. Here, $\widetilde\beta^m_z$ and $\widetilde\kappa_{z \a}$ are the antighost 
for $\widetilde\xi^m$ and for $\chi_\a$, respectively. Notice that according to this BRST charge, 
the ghosts $\widetilde\xi^m, \widetilde\chi_\a$ and $\lambda^\a$ are associated with the 
generators of the translations and of the supersymmetry transformations. 

%%%%%%%%%%%%%%%%%%%%%%%%%%%%%%%%%%%%%%%%%%%%%%%%%%%%%%%%%%%%%%%

\footatend\vfill\supereject\immediate\closeout\rfile\writestoppt
\baselineskip=14pt\centerline{{\bf References}}\bigskip{\frenchspacing%
\parindent=20pt\escapechar=` \input refs.tmp\vfill\eject}\nonfrenchspacing

\bye